\documentclass[twocolumn,linenumbers,trackchanges]{aastex62}

\usepackage{CJKutf8}

\newcommand{\cntext}[1]{\begin{CJK}{UTF8}{gbsn}#1\end{CJK}\kern-1ex}

\newcommand{\sta}{STEREO-A}
\newcommand{\sdo}{SDO}

\newcommand{\hsi}{RHESSI}
\graphicspath{{./}{figures/}}
\turnoffeditone
\nolinenumbers
\received{2020 October 29}
\revised{2021 February 5}
\accepted{2021 February 10}
\published{2021 April 8}
\submitjournal{ApJ}

\begin{document}

\title{Radio Spectral Imaging of an M8.4 Eruptive Solar Flare: Possible Evidence of a Termination Shock}

\correspondingauthor{Bin Chen}
\email{bin.chen@njit.edu}

\author[0000-0002-5431-545X]{Yingjie Luo (\cntext{骆英杰})}
\affiliation{Center for Solar-Terrestrial Research, New Jersey Institute of Technology, 323 Martin Luther King Jr Blvd., Newark, NJ 07102-1982, USA}

\author[0000-0002-0660-3350]{Bin Chen (\cntext{陈彬})}
\affiliation{Center for Solar-Terrestrial Research, New Jersey Institute of Technology, 323 Martin Luther King Jr Blvd., Newark, NJ 07102-1982, USA}

\author[0000-0003-2872-2614]{Sijie Yu (\cntext{余思捷})}
\affiliation{Center for Solar-Terrestrial Research, New Jersey Institute of Technology, 323 Martin Luther King Jr Blvd., Newark, NJ 07102-1982, USA}

\author[0000-0002-0713-0604]{T. S. Bastian}
\affiliation{National Radio Astronomy Observatory, 520 Edgemont Road,
Charlottesville, VA 22903-2475, USA}

\author[0000-0002-2002-9180]{Säm Krucker}
\affiliation{University of Applied Sciences and Arts Northwestern Switzerland, 5210 Windisch, Switzerland}
\affiliation{Space Sciences Laboratory, University of California, Berkeley, CA 94720-7450, USA}

\begin{abstract}

Solar flare termination shocks have been suggested as one of the viable mechanisms for accelerating electrons and ions to high energies. Observational evidence of such shocks, however, remains rare. Using radio dynamic spectroscopic imaging of a long-duration C1.9 flare obtained by the Karl G. Jansky Very Large Array (VLA), Chen et al.  suggested that a type of coherent radio bursts, referred to as ``stochastic spike bursts,'' were radio signatures of nonthermal electrons interacting with myriad density fluctuations at the front of a flare termination shock. Here we report another stochastic spike burst event recorded during the extended energy release phase of a long-duration M8.4-class eruptive flare on 2012 March 10. VLA radio spectroscopic imaging of the spikes \edit1{in 1.0--1.6 GHz } shows that, similar to the case of Chen et al., the burst centroids form an extended\edit1{, $\sim$10$''$-long } structure in the corona. By combining extreme-ultraviolet imaging observations of the flare from two vantage points with hard X-ray and ultraviolet observations of the flare ribbon brightenings, we reconstruct the flare arcade in three dimensions. The results show that the spike source is located \edit1{at $\sim$60 Mm } above the flare arcade, where a diffuse supra-arcade fan and multitudes of plasma downflows are present. Although the flare arcade and ribbons seen during the impulsive phase do not allow us to clearly understand how the observed spike source location is connected to the flare geometry, the cooling flare arcade observed 2 hr later suggests that the spikes are located in the above-the-loop-top region, where a termination shock presumably forms.
\nolinenumbers
\end{abstract}


\section{Introduction} \label{sec:intro}

One of the core questions in the physics of the solar flares is how a large amount of magnetic energy is converted into other forms of energy, particularly the kinetic energy in flare-accelerated particles. Although magnetic reconnection has been widely accepted as the mechanism for releasing the magnetic energy, the physical mechanisms responsible for accelerating the charged particles to nonthermal energies remain uncertain \citep[see, e.g., reviews by][]{1997JGR...10214631M, 2011SSRv..159..357Z}. Among others, shocks are believed to be an important particle accelerator in many astrophysical and space plasma environments \citep[e.g.,][]{1978ApJ...221L..29B,2008ARA&A..46...89R,2012SSRv..173....5B}. Of particular interest in the context of solar flares is the solar flare termination shock, produced by supermagnetosonic reconnection outflows impinging on the top of the newly formed flare arcades. Theoretical studies have shown that they could efficiently accelerate electrons and ions to high energies with a power-law spectral shape \citep{1997ApJ...485..859S,1998ApJ...495L..67T,2009A&A...494..669M,2009A&A...494..677W,2012ApJ...753...28G,2013ApJ...769...22L,2013PhRvL.110e1101N,2013ApJ...765..147P,2019ApJ...887L..37K,2020ApJ...905L..16K}.

Although solar flare termination shocks have been shown to exist in analytical calculations and numerical experiments \citep{1986ApJ...302L..67F,1986ApJ...305..553F,1988SoPh..117...97F,2009ApJ...701..348S,2011PhPl...18i2902W,2015Sci...350.1238C,2015ApJ...805..135T,2016ApJ...823..150T,2017ApJ...848..102T,2018ApJ...869..116S,2019MNRAS.489.3183C} and have been frequently depicted in schematics of the standard solar flare model \citep{1994Natur.371..495M,1995ApJ...451L..83S, 1996ApJ...466.1054M,1996ApJ...459..330F,2000JGR...105.2375L}, observational evidence for these shocks remains relatively rare. There have been only a handful of reports on possible signatures of flare termination shocks at radio, ultraviolet, and X-ray wavelengths \citep{1994Natur.371..495M,2002A&A...384..273A,2004ApJ...615..526A,2009A&A...494..669M, 2009A&A...494..677W,2015Sci...350.1238C,2019ApJ...884...63C,2017ApJ...846L..12G,2018ApJ...865..161P,2019MNRAS.489.3183C,2019ApJ...885...90T,2020ApJ...901...65F}. Here we refer interested readers to \citet{2019ApJ...884...63C} and references therein for more in-depth discussions regarding observational signatures of flare termination shocks, as well as possible reasons for the scarcity of these observations.  

Using radio spectroscopic imaging observations obtained by the Karl G. Jansky Very Large Array (VLA; \citealt{2011ApJ...739L...1P}) at 1--2 GHz, \cite[][hereafter Chen15]{2015Sci...350.1238C} mapped the detailed morphology and dynamics of a termination shock in a GOES-class C1.9 eruptive flare and located it at the ending point of fast plasma outflows above a loop-top hard X-ray (HXR) source. The radio signature of the termination shock manifests itself in the time--frequency domain (i.e., radio ``dynamic spectrum'') as myriad decimetric stochastic spike bursts. Each spike burst is extremely short-lived (below the instrument time resolution of 50 ms; possibly down to several milliseconds) and has a narrow frequency bandwidth $\delta \nu/\nu$ (a few percent). Chen15 interpreted the emission for the spikes as linear mode conversion of Langmuir waves excited by accelerated nonthermal electrons as they interact with small-scale density fluctuations at the shock front. Therefore, the observed centroid location of each spike burst at a central frequency $\nu$ marks the site of the corresponding density fluctuation with a mean density of \edit1{$n_e\approx ([\nu/\mathrm{Hz}]/8980)^{2}$ cm$^{-3}$}. The narrow bandwidth is determined by the amplitude level of the density fluctuation $\delta \nu/\nu \approx \frac{1}{2}\delta n_e/n_e$ \edit1{\citep{1993A&A...274..487C,2015Sci...350.1238C,2018NatCo...9..146K}}. The short temporal scale is related to the small spatial scale of the density fluctuations and the impulsive nature of the plasma radiation process. Since each density fluctuation on the shock front may have a different mean density, the ensemble of the many spike source centroids that sample a range of different $n_e$ values (for spike bursts observed in 1.0--1.8 GHz owing to fundamental plasma radiation, the density range is (1.2--4.0)$\times10^{10}$ cm$^{-3}$), in turn, outlines the detailed morphology of the shock front at any given moment. Chen15 further showed that the disruption of the shock front coincides with the reduction of both of the nonthermal radio and HXR flux, implying the role of the termination shock in accelerating electrons to at least tens of keV. 

We note that the decimetric stochastic spike bursts reported in Chen15 may or may not be the same phenomenon as the so-called ``millisecond spike bursts'' in the literature \citep[e.g.,][]{1978Natur.275..520S,1982A&A...109..305B,2002A&A...383..678B,1986SoPh..104...99B,1991ApJ...369..255G,1991A&A...251..285G,1995A&A...302..551K,1995A&A...303..249A,1996SoPh..168..375K,2001A&A...371..333P,2003ApJ...593..571F,2009A&A...499L..33B,2011ApJ...743..145C}. In fact, they appear to share some similarities with the fine structures of type II radio bursts produced by propagating coronal shocks, when high-resolution dynamic spectral data are available \citep[][hereafter Chen19]{2019ApJ...884...63C}. 
In all cases, we caution that one must be very careful when attributing an observed radio burst phenomenon with a certain appearance in the radio dynamic spectrum to a given physical interpretation, particularly when imaging data are not available. 

More recently, Chen19 studied the split-band feature of the same event reported in Chen15 with detailed dynamic spectroscopic imaging observations. 
They found that the low-frequency branch of the radio spike bursts is located slightly above its high-frequency counterpart by $\sim$0.8 Mm. Such a phenomenon \edit1{is consistent with} the scenario in which the split-band features are due to radio emission from the upstream and downstream sides of the shock front, respectively. \edit1{This interpretation is in line with recent radio imaging results for split-band features observed in type II radio bursts \citep{2012A&A...547A...6Z, 2018ApJ...868...79C,2018A&A...615A..89Z}, yet their origin is still under debate.} In this scenario, the frequency ratio between the high- and low-frequency split-band $\nu_{\rm HF}/\nu_{\rm LF}$ is determined by the density compression $n_2/n_1$ across the shock front. The shock Mach number $M$ is found to be up to $\sim$2.0, which is generally consistent with earlier predictions in analytical calculations and numerical experiments \citep{1986ApJ...305..553F,1988SoPh..117...97F,2009ApJ...701..348S,2016ApJ...823..150T,2018ApJ...869..116S}.

Here we report another decimetric stochastic spike burst event recorded by the VLA. The event was associated with a much stronger, GOES-class M8.4 eruptive flare from NOAA active region (AR) 11429 on 2012 March 10, 7 days after the C1.9 limb event reported in the earlier studies \citep{2014ApJ...794..149C,2015Sci...350.1238C,2019ApJ...884...63C}. During this period, this AR rotated from the east limb location on 2012 March 3 to the western hemisphere (N17W24), providing a unique against-the-disk view of the AR from multiple instruments. Using the same radio spectroscopic imaging technique, we derive the centroid location of the spike bursts at each time and frequency. The ensemble of the burst centroids again forms an elongated structure above the flare arcades in a diffuse supra-arcade fan (SAF) structure where plasma downflows are present. The same SAF and downflows, fortuitously, were observed by the Extreme Ultra-Violet Imager (EUVI; \citealt{2004SPIE.5171..111W}) aboard one of the Solar Terrestrial Relations Observatory satellites (STEREO; \citealt{2008SSRv..136....5K}) from a limb-view perspective. Based on 3D reconstruction of the flare arcades, we argue that the observed location and morphology of the stochastic spike bursts are consistent with the termination shock interpretation proposed in Chen15.

\begin{figure}[!ht]
\epsscale{1.2}
\plotone{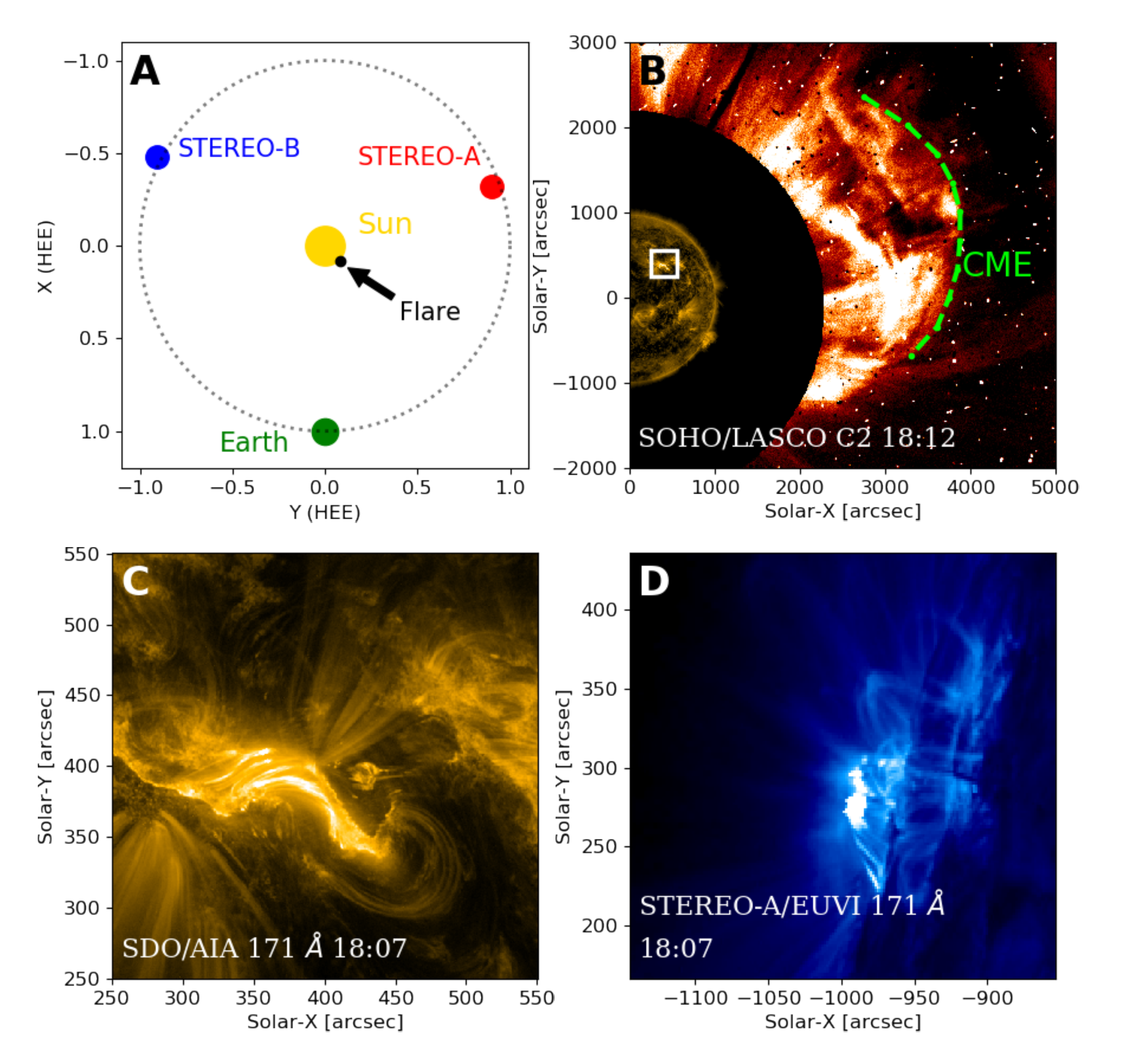}
\caption{Overview of the SOL2012-03-10T17:14:40L280C090 M8.4 eruptive solar flare event. (a) Relative location of the SDO (at a low Earth orbit) and the two STEREO satellites with regard to the flare site on 2012 March 10. (b) White-light CME observed by the SOHO/LASCO C2 coronagraph. (c) Enlarged disk view of this event from SDO/AIA 171 \AA. The flare site is marked by a white box in panel (b). (d) Limb view of this event from STEREO-A/EUVI 171 \AA.  \label{fig:overview}}
\end{figure}

\begin{figure*}
\epsscale{1.1}
\plotone{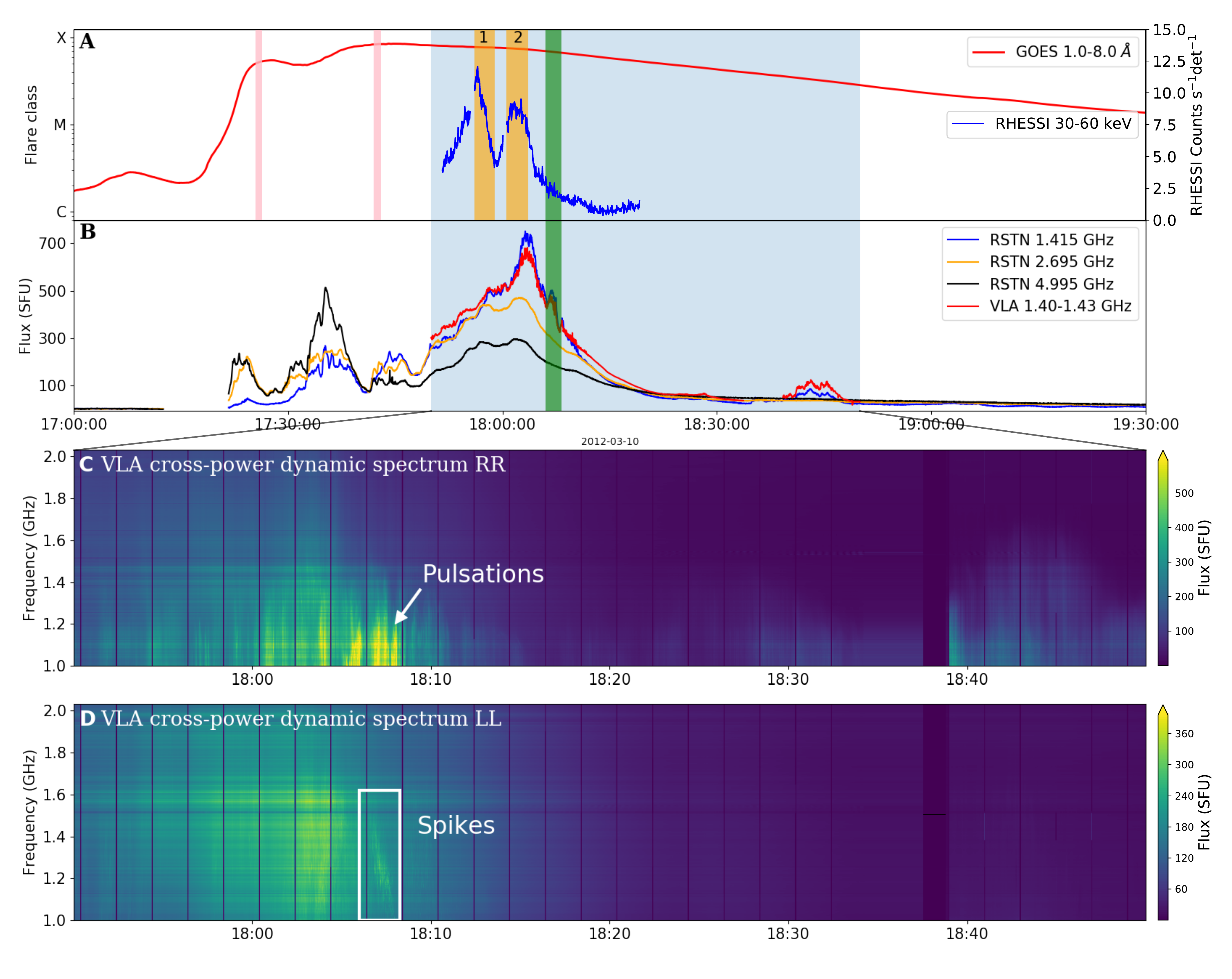}
\caption{Radio and X-ray light curves of the SOL2012-03-10T17:14:40L280C090 M8.4 eruptive solar flare event. (a) RHESSI 30--60 keV (blue) and GOES 1.0--8.0 \AA\ (red) light curves. Two pink vertical lines represent the two SXR peaks at $\sim$17:26 UT and $\sim$17:42 UT, respectively. The orange shadowed area represents the two time intervals (marked with ``1'' and ``2,'' 17:56:08--17:58:54 UT for interval 1, 18:00:35--18:03:33 UT for interval 2) used in producing the RHESSI images shown in Figs. \ref{fig:gsfit} and \ref{fig:loop}. The green shaded area represents the period when the spike bursts are present. (b) Light curves from the RSTN pre-flare background-subtracted total solar flux at 1.415 GHz (blue), RSTN 2.695 GHz (orange), RSTN 4.995 (black),  and VLA 1.40-1.43 GHz (red). (c,d) VLA radio cross-power dynamic spectra in RCP and LCP, respectively. The VLA cross-power dynamic spectra and light curve are obtained from a short baseline with a length of 181 m. The duration of the dynamic spectra is indicated by the light-blue shaded region in panels (a)--(b). The spike bursts of interest are marked by a white box in panel (d). \label{fig:lightcurve}}
\end{figure*}

After a brief description of the VLA in Section \ref{sec:vla} and an overview of the flare context in Section \ref{sec:overview}, we discuss VLA radio imaging spectroscopic observations in Section \ref{sec:radio}. In Section \ref{sec:3d}, we present 3D reconstruction of the flare arcades observed in radio, EUV, and X-rays to elucidate the location of the spike bursts above the loop tops. We also show evidence of the presence of an SAF and downflows in the close vicinity of the spike bursts. We discuss and briefly summarize our findings in Section \ref{sec:discussion}.

\section{Observations} \label{sec:observations}

\subsection{Karl G. Jansky Very Large Array} \label{sec:vla}
The VLA is a general-purpose radio telescope array that completed a major upgrade in 2011 \citep{2011ApJ...739L...1P}. The full VLA array consists of 27 antennas, each of which has a diameter of 25 m. The VLA was partially commissioned in late 2011 to enable solar observing with broadband dynamic spectroscopic imaging \citep{2013ApJ...763L..21C}. It allows radio images to be made at $>$1000 simultaneous frequency channels at an ultrahigh time cadence of up to 10 ms for each circular polarization. At the time of this study, the VLA was commissioned for solar observing at its L (1--2 GHz), S (2--4 GHz), and C (4--8 GHz) bands. Recently, support for solar observing was expanded to also include the P (0.23--0.47 GHz), X (8--12 GHz), and Ku (12--18 GHz) bands. Recent studies with the VLA have demonstrated its unique power in utilizing a variety of coherent radio bursts, including type III radio bursts, fiber bursts, and spike bursts, and incoherent gyrosynchrotron bursts to trace flare-accelerated electrons in the close vicinity of the energy release site \citep{2013ApJ...763L..21C,2014ApJ...794..149C,2015Sci...350.1238C,2018ApJ...866...62C,2019ApJ...884...63C,2017ApJ...848...77W,2019ApJ...872...71Y,2020ApJ...904...94S}.

\subsection{Event Overview} \label{sec:overview}

The event under study is a GOES-class M8.4 eruptive flare that occurred on 2012 March 10, located in AR 11429. The standard identification ID for this event is SOL2012-03-10T17:14:40L280C090 following the IAU convention \citep{2010SoPh..263....1L}.

This AR is a highly flare-productive region, which hosted more than 10 $>$M-class flares (including three X-class flares) and coronal mass ejections (CMEs) from 2012 March 2 to 17. Several studies reported the magnetic structure and evolution of this AR, as well as their relation to the flare/CME activities \citep{2014SoPh..289.2957E,2015ApJ...809...34C,2017RAA....17...81Z}. Successive eruptions and white-light CMEs from this AR were reported spanning almost the entire duration when the AR was visible from Earth  \citep{2014ApJ...794..149C,2014ApJ...788L..28L,2014ApJ...791...84W,2015ApJ...809...34C,2018ApJ...860...35D,2020ApJ...901...40D}, some of which were accompanied by shock waves \citep{2014ApJ...791..115M}, solar energetic particle (SEP) events \citep{2016ApJ...821...31K}, and high-energy gamma-ray emission \citep{2014ApJ...789...20A}. This AR is also of special interest for space-weather-related studies. In particular, the two X-class flares on 2012 March 7, accompanied by two ultrafast ($>$2,000 km~s$^{-1}$) CMEs, led to one of the largest geomagnetic storms of solar cycle 24 \citep{2016ApJ...817...14P}.

The M8.4-class eruptive flare event under study was observed by the Atmospheric Imaging Assembly (AIA; \citealt{2012SoPh..275...17L}) aboard the Solar Dynamics Observatory (SDO; \citealt{2012SoPh..275....3P}) at multiple EUV bands against the disk (Figure \ref{fig:overview}(c)). One of the STEREO satellites (\sta) also observed this event (Figure \ref{fig:overview}(d)). The spacecraft was located at 109.5$^{\circ}$ west from Earth at the time of the observation (Figure~\ref{fig:overview}(a)), and had an almost perfect limb view of this event (Figure~\ref{fig:overview}(d)). The two EUV imagers combined thus provided a unique view of the flare event from two different vantage points at the same time. 

By utilizing both \sdo/AIA and \sta/EUVI observations, \citet{2018ApJ...860...35D} studied the 2012 March 10 M8.4 event in detail. They concluded that the event was driven by a ``compound'' eruption of two filaments with a ``double-decker'' configuration. In Section \ref{sec:3d}, we will discuss our 3D reconstruction of the flare loops within the eruption context set forth in their study.

The GOES 1--8 \AA\ soft X-ray (SXR) flux started to increase at $\sim$17:15 UT. The light curve had two peaks, one at $\sim$17:26 UT and another at $\sim$17:42 UT (Figure \ref{fig:lightcurve}(a); see also \citealt{2018ApJ...860...35D}). This flare is located at 17$^{\circ}$ north and 24$^{\circ}$ west (in heliographic longitude and latitude, respectively, or N17W24) on the solar disk. It was associated with a fast white-light halo CME (Figure~\ref{fig:overview}(b)) whose speed exceeded 1200 km s$^{-1}$. 

VLA observed this flare event from 17:46:18 UT to 21:46:46 UT in its C configuration with 15 antennas. The longest baseline is 2669 m (which corresponds to an angular resolution of $\sim$19$''$ at 1.5 GHz). The observation was made in both right- and left-hand circular polarizations (RCP and LCP) in the 1--2 GHz \textit{L} band. The frequency band was divided into eight independent spectral windows, each of which had 128 1 MHz wide spectral channels. Therefore, it provided spectroscopic imaging at a total of 1024 independent frequency channels simultaneously with a time cadence of 50 ms. Flux, bandpass, and delay calibration were performed against 3C48, and antenna gain calibration was performed against J2130+0502, both of which are standard celestial calibrator sources for VLA. Solar 20-dB attenuators are inserted in the signal path during solar scans to reduce the antenna gain, and corresponding corrections in phase and amplitude were made to the data \citep{2013ApJ...763L..21C}. 

A rich variety of decimetric radio bursts were observed by the VLA during the entire duration of its 4 hr observing, which included broadband continuum emission, zebra-pattern bursts, fiber bursts, pulsations, and spike bursts. Figures~\ref{fig:lightcurve}(c) and (d) show the VLA cross-power radio dynamic spectra (from one short baseline with a length of 181 m) from 17:50 UT to 18:50 UT in RCP and LCP, respectively. The stochastic spike bursts were observed at around 18:07 UT during the extended energy release phase, $\sim$5 minutes after the HXR peak at 18:02 UT (green vertical bar in Figures \ref{fig:overview}(a) and (b) and white box in Figure \ref{fig:overview}(d)). In the following section, we will investigate these spike bursts in detail by utilizing radio dynamic spectroscopic imaging, and we suggest that they are likely associated with a flare termination shock in this event, similar to that reported in Chen15. 

\subsection{Radio Imaging Spectroscopy} \label{sec:radio}

\begin{figure*}
\epsscale{1.1}
\plotone{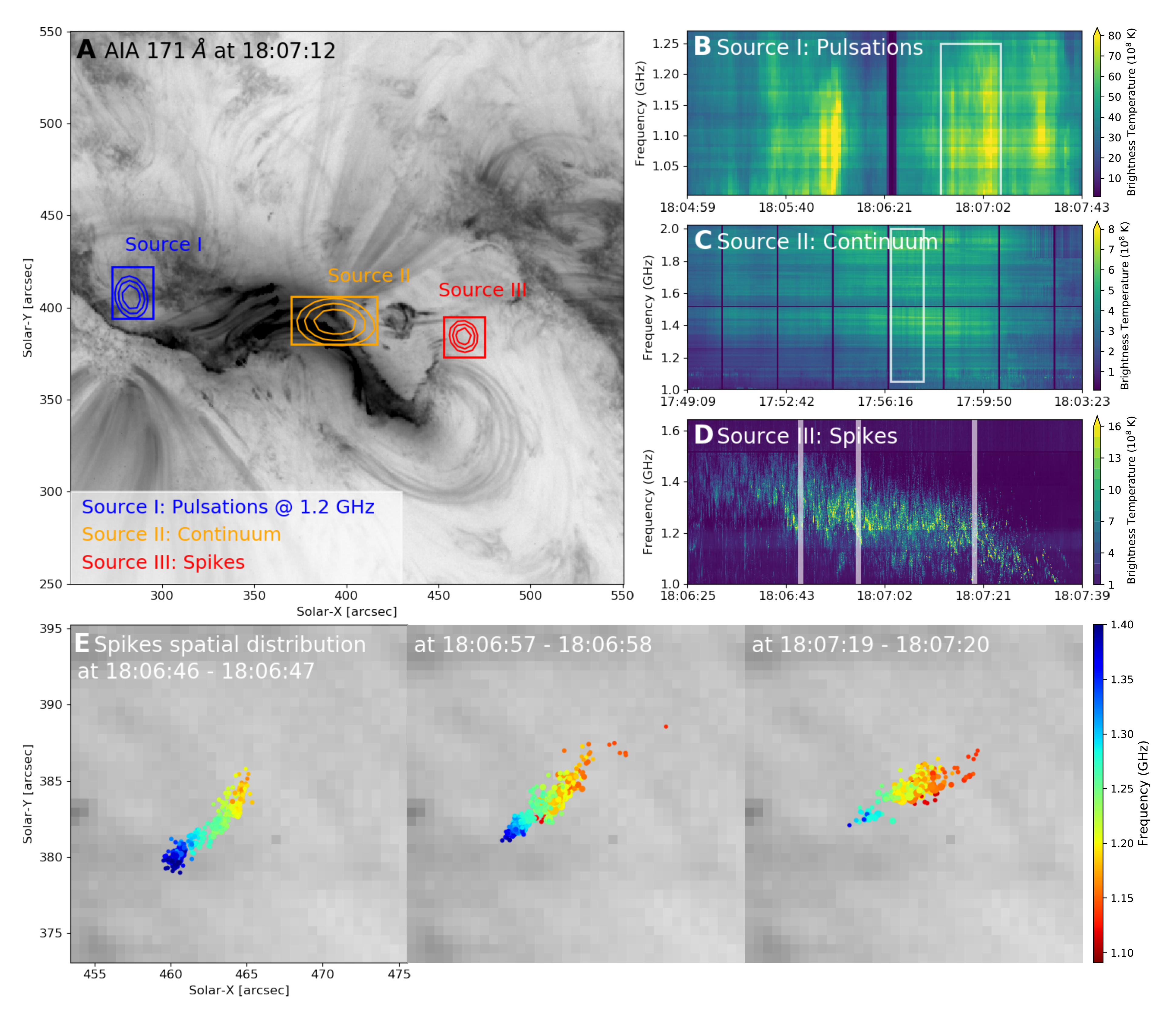}
\caption{VLA dynamic spectroscopic imaging of different radio sources during the extended energy release phase of the flare. (a) Radio images for the three sources near the time of the spike bursts shown as contours in different colors (contour levels are 70\%, 80\%, and 90\% of the maximum). Background (inverted gray-scale) image is SDO/AIA 171 \AA. (b--d) Spatially resolved vector radio dynamic spectra associated with the pulsation source located near the east flare ribbon (Source I; shown in RCP), the loop-top continuum source (Source II; shown in Stokes \textit{I}) and the stochastic spike source (Source III; shown in LCP), respectively. White boxes in panels (b) and (c) indicate the time and frequency range of the corresponding radio images (Sources I and II) shown in panel (a). For Source III, the image shown is made by taking a 70 s integration (18:06:25--18:07:35 UT) in 1.0--1.5 GHz. (e) Distributions of the centroid locations of all the spike bursts (Source III) within each of the three selected 1-s periods (indicated by vertical shades in panel (d)). Red to blue indicate increasing frequencies. The field of view is the same as the red box in panel (a). \edit1{An accompanying animation is available online. The left panel of the animation is the vector dynamic spectrum identical to panel (d), with a vertical strip indicating the corresponding time. The right panel of the animation shows the ensemble of the spike centroids from 18:06:25 to 18:07:25 UT similar to the snapshots in panel (e). The duration of the animation is 7 s.}}\label{fig:radio}
\end{figure*}

Around the time of interest when the spike bursts are present ($\sim$18:07 UT), radio imaging in our observing frequency range shows three distinct sources (Figure~\ref{fig:radio}(a)). One source is located close to the east flare ribbon (Source I; blue contours). Another source is located at the top of the flare arcades (Source II; orange contours). A third source is located $\sim$70$''$ away from the loop-top source toward the west (Source III; red contours). VLA's capability of radio dynamic spectroscopy imaging allows us to image every pixel in the time--frequency domain, resulting in a time series of 3D spectral image cubes (two in space, one in frequency) in each polarization product. Such an essentially 4D spectrotemporal image cube can be used to generate a spatially resolved, or ``vector'' dynamic spectrum from any selected spatial region of interest, thereby elucidating the spectrotemporal properties intrinsic to the given source \citep[e.g.,][]{2015Sci...350.1238C,2018ApJ...866...62C,2017SoPh..292..168M,2019ApJ...872...71Y}. We employ this technique to produce a vector dynamic spectrum for each of the three radio sources, shown in Figures \ref{fig:radio}(b), (c), and (d), respectively. The vector dynamic spectra for the three sources are strikingly different: Source I displays highly polarized bright pulsating signatures with many complex fine structures, with a peak brightness temperature ($T_B$) of over $10^{10}$ K. The pulsation source is possibly due to coherent radiation by trapped energetic electrons \citep[see, e.g.,][]{2019NatCo..10.2276C}, which may be the topic of a future study. Source II at the loop top, in contrast, is a weakly polarized, continuum-like emission, which has a brightness temperature that increases in frequency to about $10^9$ K. Source III, which is the focus of the current study, shows a large group of short-lived, narrow-bandwidth spike bursts. The dynamic spectral features of this source closely resemble the stochastic spike burst event reported in Chen15, who attributed them to a radio signature of emission at myriad density fluctuations at or close to the flare termination shock front. The spike bursts are highly polarized in LCP and have a peak $T_B$ of over $10^{9}$ K. 

\subsubsection{Loop-top Continuum Source}

\begin{figure*}[!ht]
\epsscale{1.1}
\plotone{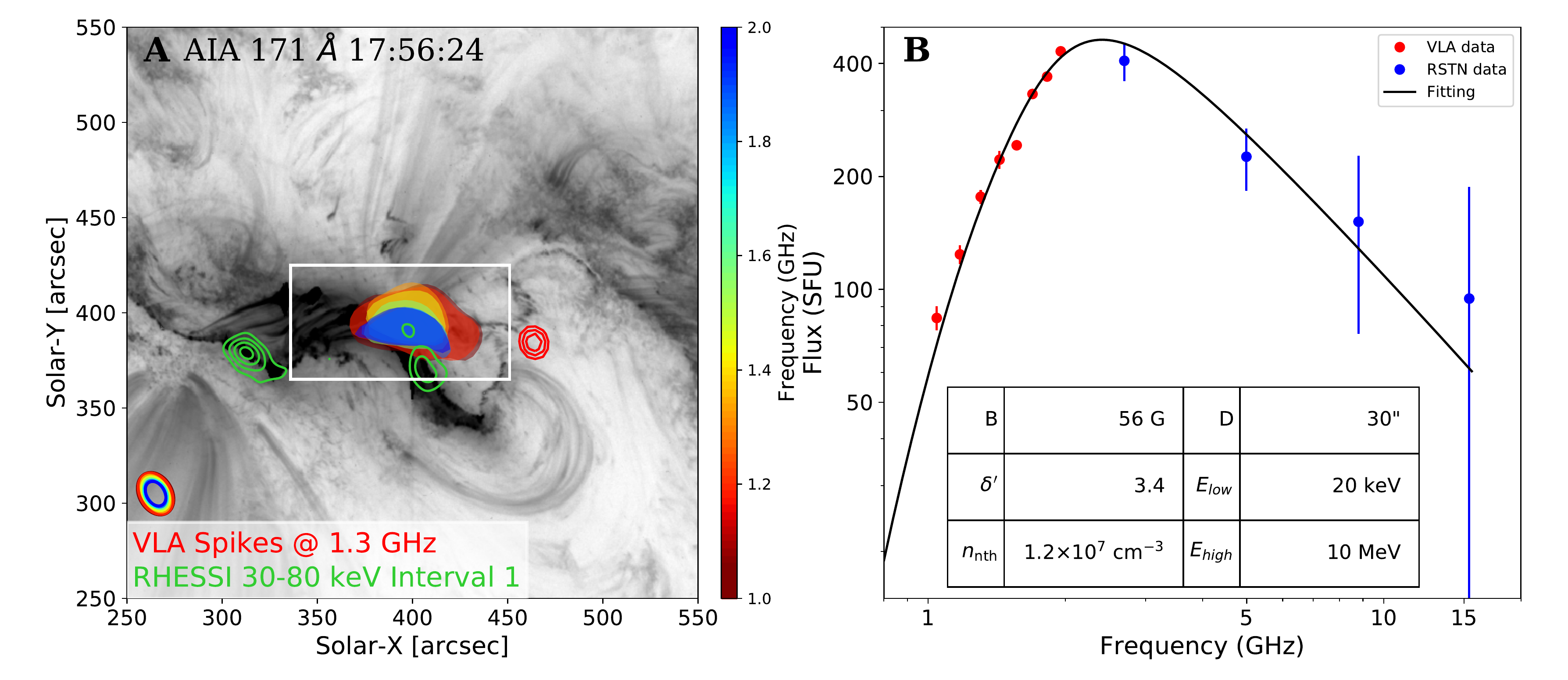}
\caption{Imaging spectroscopy of the loop-top radio continuum source (Source II in Fig. \ref{fig:radio}) and gyrosynchrotron fitting results. (a) Spectral-resolved contours of the loop-top radio continuum source at each spectral window (50\% of the maximum brightness temperature at its frequency in the loop-top area). Color from red to blue indicates increasing frequency (see color bar on the right). The corresponding FHWM beam sizes are shown in the lower left corner. Green contours are RHESSI 30--80 keV HXR footpoint sources (HXR interval 1 in Figure \ref{fig:lightcurve}(a), 17:56:08--17:58:54 UT) that coincide with the double flare ribbons. (b) Radio flux density spectrum of the loop-top radio continuum source and gyrosynchrotron fitting results. VLA measurements, shown as red circles, are derived by summing all the fluxes within the white box in panel (a). Blue circles are total-power data observed by RSTN at 2.695, 4.995, 8.8, and 15.4 GHz. 
}\label{fig:gsfit}
\end{figure*}

The rising spectrum of the loop-top continuum source (Source II in Figure~\ref{fig:radio}) in 1--2 GHz is consistent with the spectral characteristics of the optically thick portion of gyrosynchrotron radiation at the same frequency range \citep[e.g.,][]{1982ApJ...259..350D}. To further confirm the gyrosynchrotron nature of the loop-top continuum source, we utilize concurrent total-power data obtained by the Radio Solar Telescope Network (RSTN) at eight discrete frequencies: 0.245, 0.410, 0.610, 1.415, 2.695, 4.995, 8.800, and 15.40 GHz. The RSTN fluxes for the lowest three frequencies (below 1 GHz) reached over 1000 solar flux units (SFU) at $\sim$17:57, and have very different temporal behavior from their higher-frequency counterparts. They are likely dominated by coherent emission. Also, as VLA covers the 1.415 GHz band, RSTN fluxes at the lowest four frequencies are excluded from the subsequent analysis. 

To enable direct comparison to the spatially resolved VLA observations of the loop-top source, first, we subtract the post-flare background at $\sim$19:30 from the observed RSTN flux at 17:57 to remove the contribution from the quiescent solar disk and active region. Background subtraction is also performed on VLA data at 17:57 by selecting the same post-flare time as the background in the loop-top region (white box in Figure \ref{fig:gsfit}(a)). Second, we take the assumption that the radio emission above 2 GHz is dominated by the loop-top source. This assumption is mostly valid, as VLA imaging reveals that the loop-top source plays a dominant role at above $\sim$1.4 GHz (see Figure \ref{fig:lightcurve}) during the time of interest. Uncertainties of flux values derived from VLA data are estimated by using the rms variations of a region in the image without the presence of any source. For the uncertainty estimates for RSTN-derived flux values, we take the rms variations of the measured total flux during a nonflaring time.

The resulting spectrum of the loop-top source is shown in Figure~\ref{fig:gsfit}(b). With the addition of RSTN flux at higher frequencies ($>$2 GHz), the spectrum shows features characteristic of nonthermal gyrosynchrotron radiation, with a positive and negative slope at the low- and high-frequency side (due to optically thick and thin emission, respectively) and a peak flux of $\sim$400 SFU. \edit1{Although the combined microwave spectrum does not allow us to accurately determine the source parameters owing to the assumptions taken above and the lack of spatial resolution across the spectrum, for demonstration purposes} we perform spectral fitting based on a homogeneous source model by adopting the fast gyrosynchrotron codes in \citet{2010ApJ...721.1127F} to calculate the gyrosynchrotron spectrum. \edit1{An example fit result and the associated source parameters are shown in Figure~\ref{fig:gsfit}(b)}\footnote{Source parameters listed in Figure~\ref{fig:gsfit}(b) are magnetic field strength $B$, power-law index $\delta'$ of the electron density distribution, total nonthermal electron number density $n_\mathrm{nth}$, low- and high-energy cutoff of the power-law nonthermal electron distribution $E_{\rm low}$ and $E_{\rm high}$, and column depth $D$.}.
\edit1{Although the fit parameters are subjected to the limitations noted above and are probably not unique}, our analysis demonstrates that the loop-top continuum source is most likely due to the incoherent gyrosynchrotron emission from flare-accelerated nonthermal electrons.

\subsubsection{Stochastic Spike Source}
As mentioned earlier, the vector dynamic spectral features of Source III closely resemble the stochastic spike burst event reported in Chen15. The group of spike bursts lasted for around 70 s from 18:06:25 to 18:07:35 UT in the frequency range of 1.0--1.6 GHz. It contains a total of over 10,000 individual bursts within the $\sim$70s duration. Similar to the spike bursts reported by Chen15, the bursts are extremely short-lived---most are unresolved by the time resolution of the instrument 50 ms---and have a narrow bandwidth $\delta\nu/\nu$ of only a few percent. 

\begin{figure*}[!ht]
\epsscale{1.0}
\plotone{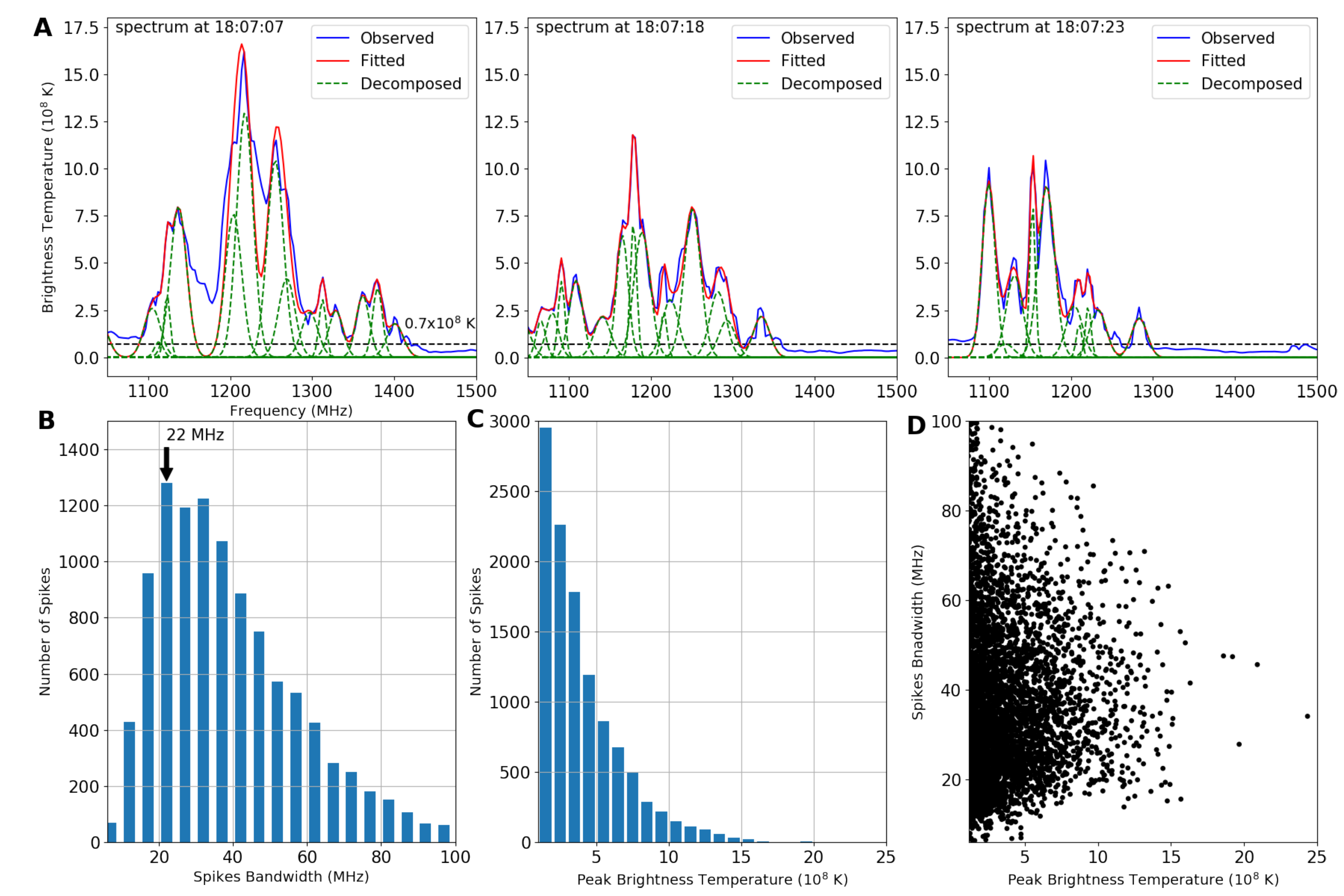}
\caption{Spectral characteristics of spikes. (a) Examples of multi-Gaussian spectral fitting results for the VLA radio spectra at three different times. Blue and red solid lines are the observed and fitted spectra. Green dashed curves are the multiple Gaussian components used in the fitting. The $T_B$ threshold adopted for the fitting is indicated by a horizontal dashed line. (b,c) Bandwidth and peak brightness temperature distribution of the spikes from the fitting results. (d) Scatter plot of the peak brightness temperature vs. bandwidth of the spikes.}\label{fig:bd-dis}
\end{figure*}

To further study the spectral properties of the spike bursts, we adopt the same technique described in Chen15 to perform a multi-Gaussian fit of the spectral profiles of the spike bursts at each time integration. To reduce the noise in the spectral domain and achieve more robust fit results, we have smoothed the spectral profiles with a 6MHz wide boxcar window. Further, in order to limit our fitting to sufficiently bright spike bursts, we have also applied a brightness threshold of $T_B^\textrm{min}=0.7\times10^8$ K, set as about 4 times of the rms variations of a region in the image without any source. The fitting range of the spike bandwidth (represented by the FWHM of the Gaussian profiles) is limited to 6--100 MHz. Typical examples of the fit results for three selected times are shown in Figure~ \ref{fig:bd-dis}(a). From the multi-Gaussian fitting results we obtain a distribution of the spike bandwidth, shown as a histogram in Figure~\ref{fig:bd-dis}(b). Similar to the bandwidth distribution of the spike burst event shown in Chen15 (their Figure S3 in Supplementary Materials), the spikes have an asymmetric distribution that peaks at 22 MHz (or a relative bandwidth $\delta\nu/\nu\approx2\%$, density fluctuation $\delta$$n_{e}$/$n_{e}$$\approx4\%$). The brightness temperature distribution is also notably similar to that shown in Chen15. We note, however, that the maximum brightness temperature is two orders of magnitude greater than that reported in Chen15. \edit1{It can be possibly attributed to the much larger flare (M8.4 vs. C1.9) in which our spike bursts are observed (see, e.g., \citealt{2005SoPh..226..121B}, who reported that coherent radio bursts are generally brighter in larger flares), yet the underlying relationship is highly complex, which involves many factors, including flare timing, shock generation, electron acceleration, and the coherent radiation processes.} Nevertheless, the very similar spectrotemporal properties to those reported in Chen15 suggest that the spike bursts are likely associated with the same emission processes. 

We further derive the centroid location of the radio source at each time and frequency pixel where the spike burst is present by using a 2D polynomial fitting technique following Chen15. The 2D polynomial fitting utilizes the 6 pixels surrounding the pixel where the maximum brightness is located to perform second-order polynomial fitting in two orthogonal directions. The location of each source centroid is found at the peak of the polynomial curves. The uncertainty of the resulting centroid position is estimated by $\sigma\approx\theta_{\rm FWHM}/({\rm S/N}\sqrt{8\ln2})$ \citep{1988ApJ...330..809R,1997PASP..109..166C,2015Sci...350.1238C,2018ApJ...866...62C}, where the signal-to-noise ratio (S/N) is the ratio of the peak flux to the rms noise of the synthesized image where a source is not present, and $\theta_{\rm FWHM}$ is the FWHM of the synthesized beam. With the $\theta_{\rm FWHM}=19.''9\times14.''6 $ at 1.6 GHz and typical S/N values greater than 40, the accuracy of our derived source centroid position is typically better than 1$''$. 

As shown in Figure~\ref{fig:radio}(e), at any given time, the derived centroid locations of the spike bursts at different frequencies form a nearly linear feature with a length of about 10$''$. This distinctive feature shows some temporal evolution and lasts for about 70 s before it diminishes at 18:07:35 UT (see Figure~\ref{fig:radio}(e) for snapshots at three different times and the associated animation). This feature shows a slight overall movement toward the southwest direction (or a movement toward the lower right corner in Figure~\ref{fig:radio}), which corresponds to a gradual overall frequency drift of the spike burst group toward lower frequencies in the dynamic spectrum (Figure \ref{fig:radio}(d)). The observed features closely resemble the findings in Chen15, who interpreted the feature as the projection of \edit1{a presumably 3D} termination shock front above the flare arcades, and the overall frequency drift as the movement of the shock front in the loop-top region with a varying plasma density. We note, however, that unlike the event reported in Chen15 and Chen19, this event does not show any sign of sudden disruption of the feature or a split-band feature. 

Although VLA radio imaging shows that the spike source is located away from the flare arcades, as expected from the termination shock scenario in which the shock front is situated above the loop tops, it is not immediately clear whether the exact location and orientation of the spike source are consistent with the termination shock interpretation. In the next two subsections, we will take advantage of the concurrent observations obtained by both \sdo/AIA and \sta/EUVI from two vantage viewing points and establish the spatial relation between the eruption, the flare arcades, and the SAF and downflow structures.

\subsection{Spike Source Location in a 3D Context} \label{sec:3d}
Unlike the stochastic spike burst event reported in Chen15, which had a limb view, the spike source under study here is observed against the disk. Therefore, the projection effect renders our interpretation for the spike source in the flare context not as straightforward as the limb event. Fortuitously, as reported by \citet{2018ApJ...860...35D}, the event was recorded simultaneously by \sdo/AIA and \sta/EUVI from two viewing perspectives: \sdo\ viewed it against the disk, and \sta\ viewed the eruption from the limb (panels (a) and (c) in Figure \ref{fig:flux_rope}, respectively). These observations give us a unique opportunity for understanding the location of the radio sources in a 3D flare context. In this subsection, we will first reconstruct the geometry of the erupting filament and flare arcades using observations from two vintage viewing perspectives. We will then place the observed spike bursts, loop-top gyrosynchrotron radio source, and SXR source in the physical context of this eruptive flare. 

\subsubsection{Erupting Filament}
\begin{figure*}[!ht]
\epsscale{1.0}
\plotone{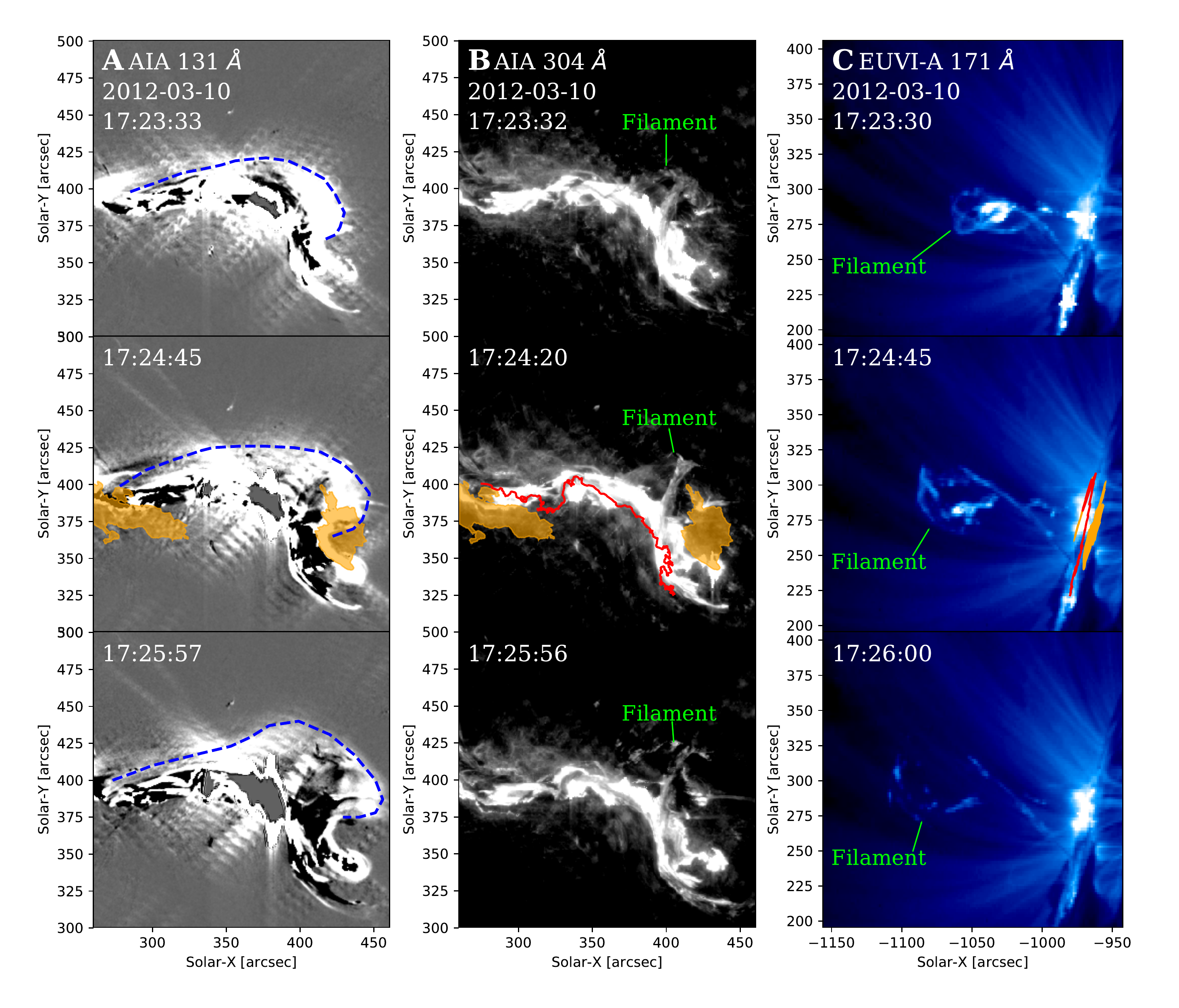}
\caption{Erupting filament structure observed by (a) \sdo/AIA 131 , (b) 304 \AA\, and (c) STEREO-A/EUVI 171 \AA\  from two viewing perspectives. (a) \sdo/AIA 131 \AA\ running-difference images, with the blue dashed curves representing the eruption front. \edit1{The PIL is shown as a red curve in panels (b) and (c). Orange patches represent the coronal dimming region detected from the \sdo/AIA 211 \AA\ base-difference image (post-flare vs. pre-flare), which coincides with the ends of the eruption on opposite sides of the PIL. The orange patches in panel (c) are the same dimming regions transformed to \sta/EUVI's perspective.}
\label{fig:flux_rope}}
\end{figure*}

In \citet{2018ApJ...860...35D}, by using both \sdo/AIA and \sta/EUVI observations, the authors suggested that the event was driven by two preexisting filaments with a ``double-decker'' configuration. As observed by \sdo/AIA, the filaments were formed prior to the flare along the primary polarity inversion line (PIL\edit1{; red curve in Figure~\ref{fig:flux_rope}}) with a northeast--southwest orientation. The double-decker filaments erupted successively $\sim$12 minutes apart prior to the second GOES SXR derivative peak at $\sim$17:37 UT. After $\sim$17:30 UT, the two erupting filaments merged together and became nearly indistinguishable from each other. The erupting filament, seen in \sdo/AIA 304 \AA\, has a writhed shape and directs toward the northwest (Figure~\ref{fig:flux_rope}(b)). The eruption is less evident in \sdo/AIA 131 \AA\ images but can be distinguished using the running-difference imaging technique (Figure~\ref{fig:flux_rope}(A)). This passband is sensitive to a much higher temperature ($\sim$10 MK) associated with the \ion{Fe}{21} line \citep{2010A&A...521A..21O}, which is presumably emitted by the hot envelope of the magnetic flux rope encompassing (and likely above) the cool filament \citep[e.g.,][]{2014ApJ...794..149C,2014ApJ...789L..35C}. Similar to \citet{2018ApJ...860...35D}, we track the eruption front seen in \sdo/AIA 131 \AA, shown as blue dashed lines in Figure~\ref{fig:flux_rope}(a). In \sta/EUVI's limb view, the erupting filament clearly displays a writhed morphology (Figure \ref{fig:flux_rope}(c)), which strongly suggests that it is a magnetic flux rope in nature \citep[e.g.,][]{2003ApJ...595L.135J,2005ApJ...630L..97T} and is consistent with the morphology of the filament viewed from \sdo/AIA. 

\edit1{The conjugate ends of the eruption coincide with two coronal dimming regions in \sdo/AIA 211 \AA\ base-difference images (difference of the post- and pre-flare intensity; orange patches in Figure~\ref{fig:flux_rope}) located at the opposite sides of the PIL. These twin coronal dimmings, often observed during the eruption phase of CME-associated events in EUV and SXR, have been interpreted as density depletions at the conjugate ends of the expelled magnetic flux rope \citep[e.g.,][]{1997ApJ...491L..55S,1999ApJ...520L.139Z,2011SoPh..273..125M,2016ApJ...825...37C}. Their presence further supports the flux rope nature of the erupting filament.} 

\subsubsection{Flare Arcades}
\begin{figure*}[!ht]
\epsscale{1.2}
\plotone{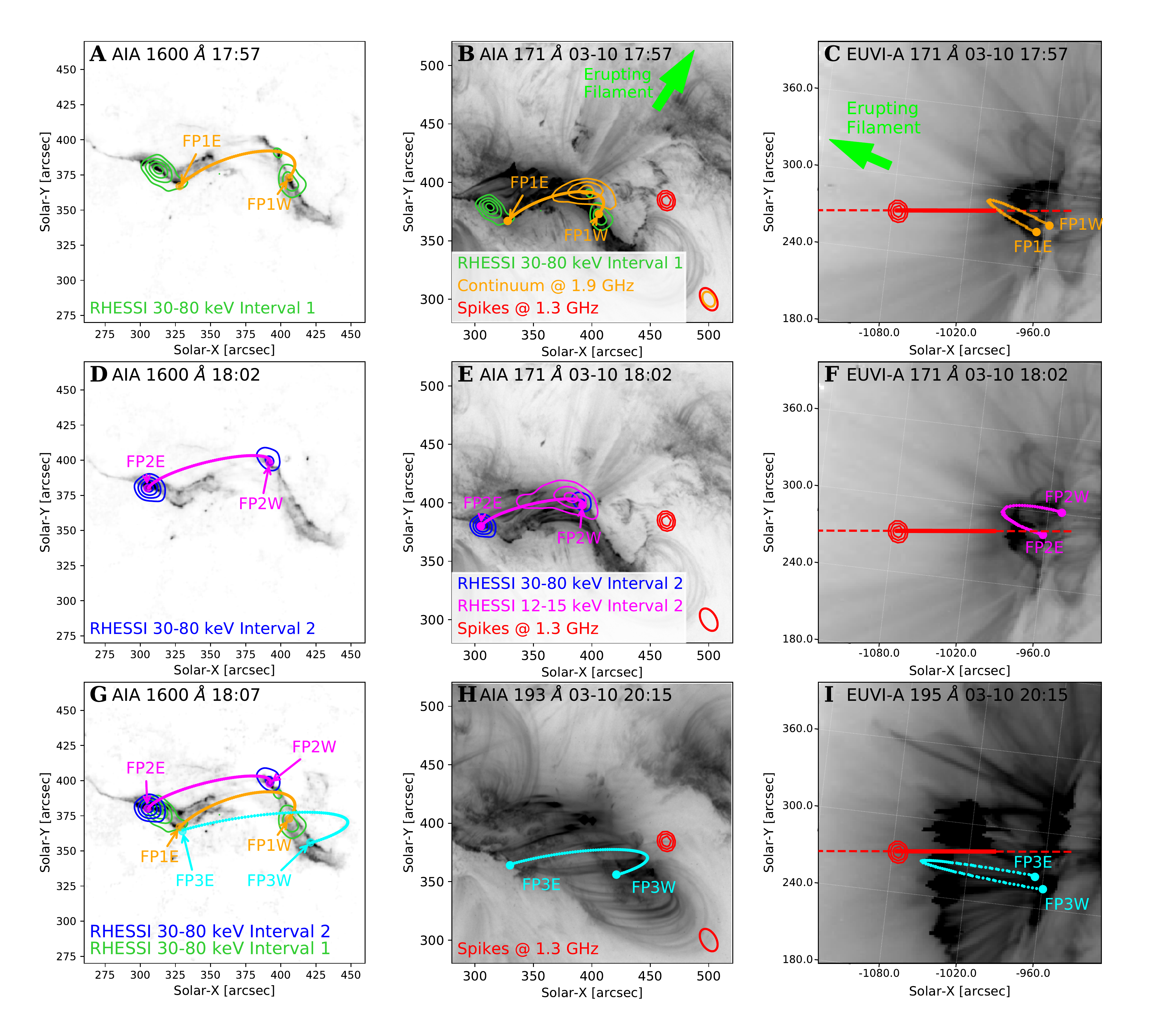}
\caption{3D reconstruction of flare arcades. (a, d) \hsi\ 30--80 keV footpoint sources at intervals 1 and 2 (17:56:08 UT-- 17:58:54 UT and 18:00:35 UT--18:03:33 UT, respectively) and the concurrent \sdo/AIA 1600 \AA\ images shown in inverted gray scale. (b, c) 3D reconstructed loop (thick orange curve) that connects the footpoint sources and the VLA loop-top radio continuum source (orange contours) in \sdo\ and \sta's view. Green arrows illustrate the direction of the erupting filament (see Figure \ref{fig:flux_rope}). The spike burst is shown as red contours in panel (b). (e, f) Similar to panels (b) and (c), but instead showing the 3D reconstructed loop (thick magenta curve) based on the SXR arcade (magenta contours) and the HXR footpoints (blue contours) in interval 2. The LOS of an Earth-based observer passing the spike source in \sta's view is shown as a red dashed line. \edit1{The solid portion shows the range of the possible spike source location above the apex of the transformed SXR arcade, with its best estimate shown as the red contours}. (g) Similar to panels (a) and (d), but includes a reconstructed post-flare arcade in \sdo/AIA 193 \AA\ and \sta/EUVI 195 \AA\ images at about 2 hr later (cyan curve). The latter is shown in panels (h) and (i). The size of the synthesized beam is shown in lower right corner.
\label{fig:loop}}
\end{figure*}

The appearance of the flare arcades, as viewed by \sdo/AIA and \sta/EUVI, is consistent with the geometry of the erupting filament (Figure~\ref{fig:loop}). In \sdo/AIA's view, the post-flare arcades are distributed from northeast to southwest, rooted at either side of the pre-flare filament (Figure~\ref{fig:loop}, middle column). In \sta/EUVI's limb view, the same post-flare arcade system is projected to distribute in the north--south direction along the limb, which is consistent with the geometry of the filament that erupted early on (Figure \ref{fig:loop}, right column). According to the standard picture of eruptive solar flares in three dimensions \citep{2012A&A...543A.110A,2013A&A...549A..66A,2013A&A...555A..77J,2014ApJ...788...60J,2015SoPh..290.3425J}, a large-scale reconnection current sheet is present below the erupting filament and above the post-flare arcade, driving the flare energy release and particle acceleration \citep[see, e.g., a recent study by][]{2020ApJ...895L..50C}.

\begin{figure*}[!ht]
\epsscale{1.2}
\plotone{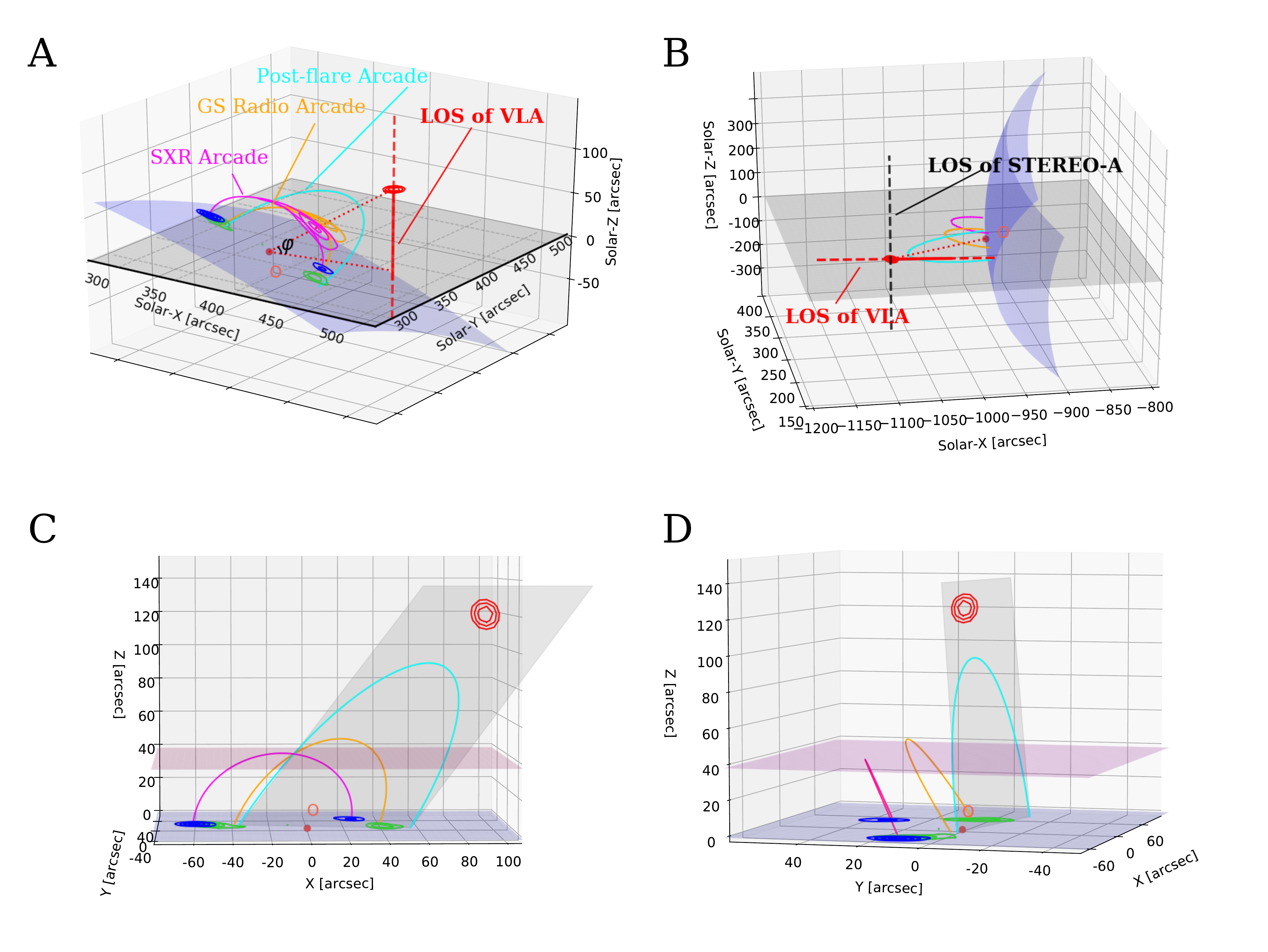}
\caption{3D view of the reconstructed loops. (a, b) Reconstructed loops in \sdo's\ and \sta's viewing perspective, respectively. They are identical to those in Figure \ref{fig:loop}: magenta, orange, and cyan curves are for the SXR arcade, the GS radio arcade, and the post-flare arcade at 20:15 UT, respectively. The \textit{X--Y} plane is the plane of the sky in each perspective. The Z-axis is the direction orthogonal to the \textit{X--Y} plane, i.e., the LOS of each viewing perspective. The blue curved surface represents the solar surface. The projection angle $\varphi$ discussed in the text is defined as the angle between the line connecting the \edit1{best estimate of the} spike source location (red contours) and the plane of the sky in SDO's view. (c, d): Detailed view of the spike source with respect to the three reconstructed loops in a face-on and edge-on projection. The gray surface represents the plane of the newly reconnected arcade (cyan curve). The blue shaded bottom plane is (approximately) the solar surface, with the origin of the \textit{X-, Y-}, and \textit{Z}-coordinates located at the midpoint of the radio GS arcade (orange curve) marked as ``o''. \edit1{The pink surface marks the radial height of the apex of the SXR arcade.}\label{fig:loop3d}}
\end{figure*}

During the flare energy release phase, magnetic reconnection is likely patchy and intermittent \citep[e.g.,][]{2002SSRv..101....1A}. At any given time, reconnection occurs within a localized region, forming a newly reconnected flare arcade. Downward-propagating nonthermal electrons propagate along the legs of the flare arcade and produce a pair of compact HXR sources at the conjugate footpoints of the flare arcade as they bombard the dense chromosphere. Chromospheric material heated by the electron beams and/or thermal conduction at different times forms the observed flare ribbons at (E)UV and/or white-light wavelengths, which are usually much more extended than the HXR footpoint source owing to their much slower decay \citep[see, e.g., discussions in][]{2009A&A...493..241F,2010ApJ...725..319Q}. 

In our event, \hsi\ 30--80 keV images made during the extended energy release phase (green and blue contours in Figures \ref{fig:loop}(a), (d), and (g), which correspond to the two intervals we selected for imaging shown in the light curve in Figure \ref{fig:lightcurve}(a)) reveal multiple pairs of conjugate HXR footpoint sources distributed along the double flare ribbons. The latter is clearly visible in \sdo/AIA 1600 \AA\ images (inverse gray-scale background in Figure \ref{fig:loop}). 
The location and morphology of the HXR footpoints and UV flare ribbons are consistent with the standard flare scenario discussed above. The flare arcade itself is filled with flare-heated plasma, producing thermal SXR emission (\hsi\ 12--15 keV; magenta contours in Figure~\ref{fig:loop}(e)). \edit1{X-ray spectral analysis using the \texttt{OSPEX} package available in \texttt{SolarSoft IDL} suggests a volume emission measure of $7.8\times10^{48}$ cm$^{-3}$ and a temperature of 19 MK (Appendix A).} Flare-accelerated electrons are often trapped at or above the top of the flare arcade, producing (above) loop-top HXR and/or radio sources \citep[e.g.,][]{2018ApJ...863...83G,2020NatAs...4.1140C}. In this event, as already discussed in Section~\ref{sec:radio}, such a loop-top radio source is also present, likely due to gyrosynchrotron radiation from flare-accelerated nonthermal electrons. A loop-top HXR counterpart is not detected by \hsi. We attribute such a nondetection to its limited dynamic range for imaging the presumably weak coronal source in the presence of strong footpoint HXR sources \citep[e.g.,][]{2008A&ARv..16..155K}. 

Both the RHESSI 12--15 keV SXR source and VLA gyrosynchrotron radio source display an arcade shape. We adopt a semi-ellipse 3D loop modeling approach to match the morphology of the observed SXR and radio arcade, using the HXR footpoint sources and bright UV flare ribbon kernels as their anchoring points. Free parameters in the 3D loop model include the major and minor axis of the ellipse $a$ and $b$ (with their ratio $e=a/b$), the inclination angle $\alpha$ of the ellipse plane about the plane perpendicular to the solar surface (a positive angle means a northward inclination), and the rotation angle $\theta$ of the ellipse about its center ($\theta=0^{\circ}$ corresponds to a ellipse with its vertices coinciding with the HXR footpoints or bright ribbon kernels). The relation between the separation of the loop footpoints (2$l$) in terms of $a$, $b$, and $\theta$ is
\begin{equation}
    l^2=a^2b^2/(b^2\cos^2\theta+a^2\sin^2\theta).
\end{equation} 
With $l$ fixed using the observed footpoint separation, the remaining free parameters that define the 3D semi-ellipse loop are $e=a/b$, the rotation angle $\theta$, and the inclination angle $\alpha$. We adjust these parameters until each loop model in projection (from Earth view) matches the morphology of the observed loop-top radio and SXR arcade. The resulting best-fit semi-ellipse loops are shown as the thick orange and magenta curve in Figures \ref{fig:loop}(b) and (e), respectively. The same loop models are then projected from the viewing perspective of \sta/EUVI (orange and magenta curves in Figures~\ref{fig:loop}(c) and (f). We will refer to the two model arcades in which the gyrosynchrotron radio source and loop-top SXR are situated, respectively, as the ``GS radio arcade'' and ``SXR arcade'' hereafter.

\subsubsection{Spike Source Location}\label{sec:spikes_3d}
The spike source, as viewed from Earth by the VLA, is located $\sim$70$''$ westward in projection from the top of the GS radio arcade (Figure~\ref{fig:loop}(b)). Unlike the loop-top SXR and GS radio sources that have an arcade-like shape, the spike source is unresolved (the synthesized beam is shown in all panels of Figure \ref{fig:loop} in the \edit1{center} column as an ellipse). Therefore, the same loop modeling technique described previously cannot be applied to constrain its 3D geometry. \edit1{Without further constraints, the spike source can, in principle, be located anywhere along the line of sight (LOS) of VLA, shown as a red dashed line in STEREO-A's view in Figures \ref{fig:loop} and \ref{fig:loop3d}. However, if we adopt the termination shock scenario, it is reasonable to assume that the spike source is located above the top of the flare arcade and} makes the same projection angle $\varphi$ as the loop top of the GS arcade, defined as the angle between the semiminor axis of the semi-ellipse loop model for the GS radio arcade and the plane of the sky ($\varphi\approx 43^\circ$; shown in Figure \ref{fig:loop3d}(a)). Therefore, a displacement of $d\approx98''$ from the center of the semi-ellipse loop model for the GS radio arcade (marked as ``o'' in Figure \ref{fig:loop3d}(a)) to the spike source in projection translates into an absolute distance $D=d/\cos\varphi\approx134''$. Knowing the location of the spike source in 3D, we can then derive its (radial) height above the solar surface to be $h\approx120''\approx86$ Mm. 
Adopting such a height, in \sta/EUVI's view, we mark the inferred spike source location as red contours, located well above the top of the flare arcades. The plane defined by the inferred 3D location of the spike source and the conjugate HXR footpoints is close to the plane where the GS radio arcade is situated, but it is tilted further toward solar south by $\sim$21$^\circ$. If we apply the same method on the SXR arcade to estimate the spike's location, it gives a slight difference ($\sim$5$''$) in the radial height of the spike source.
Figure \ref{fig:loop3d} visualizes the 3D spatial relation among the SXR arcade, the GS radio arcade, and the location of the spike source from different viewing perspectives. 

Our 3D reconstruction results place the \edit1{best estimate of the} spike source location at about $\sim$120$''$ above the solar surface, or $\sim$60 Mm above the apex of the flare arcade. This is generally consistent with the termination shock scenario, in which a shock front is formed at the end of reconnection outflows above newly reconnected, cusp-shaped magnetic loops (see, e.g., MHD simulations in Chen15 and \citealt{2018ApJ...869..116S}). For this event, as viewed against the disk by \sdo/AIA, these newly reconnected loops cannot be clearly distinguished owing to the lack of column depth along the LOS. However, it is expected that they will relax and cool down to background coronal temperatures due to conductive and radiative cooling during the decay phase. 

To look for signatures of loops formed around the time of the spike bursts, first, we use the formula given by \citet{1995ApJ...439.1034C} (their Equation 14(E)),
\begin{equation}
\tau_c = 0.0235\frac{L^{5/6}}{(Tn)^{1/6}},
\end{equation}
where $T$ and $n$ are the initial temperature (in K) and density (in cm$^{-3}$) of a loop with a half-length $L$ (cm), to estimate the loop cooling time. For a loop half-length of 10$^{10}$ cm (or 100 Mm; estimated using the distance from the spike source to the center of the semi-ellipse model loop for the GS radio arcade $D$), and plasma temperature of 20 MK, and density of 10$^9$--10$^{10}$ cm$^{-3}$ typical for flare-heated newly reconnected loops, the loop cooling time $\tau_c$ is estimated as $\sim$2--3 hr. Using this cooling time as a guide, in Figure \ref{fig:loop}, we show \sdo/AIA 193 \AA\ and \sta/EUVI 195 images (differentially rotated back to 18:07 UT; the passbands are sensitive to 1--2 MK coronal plasma; see, e.g., \citealt{2008SSRv..136...67H,2010A&A...521A..21O}). In these images, we find multitudes of well-defined, cool post-flare arcades. 
One representative semi-ellipse loop model that fits the observed AIA and EUVI arcade features is shown as the thick cyan curve in Figures \ref{fig:loop}(h) and (i). The apex of the loop is found to be present just below the \edit1{best-estimate location of the} spike source, which is consistent with the scenario of freshly reconnected loops located just beneath a termination shock front. We stress that the semi-elliptical EUV loops shown here may only represent the already-relaxed (and cooled-down) state of the reconnected loops after their shrinkage due to the magnetic tension force \citep[e.g.,][]{2008ApJ...675..868R}. The newly reconnected loops themselves are usually observed to show a cusp shape when viewed from the side \citep[e.g.,][]{1992PASJ...44L..63T, 1996ApJ...456..840T}, which are not visible in this event owing to the unfavorable viewing perspective.   

\subsubsection{Super-arcade Fan and Downflows} \label{sec:sad}

\begin{figure}
\plotone{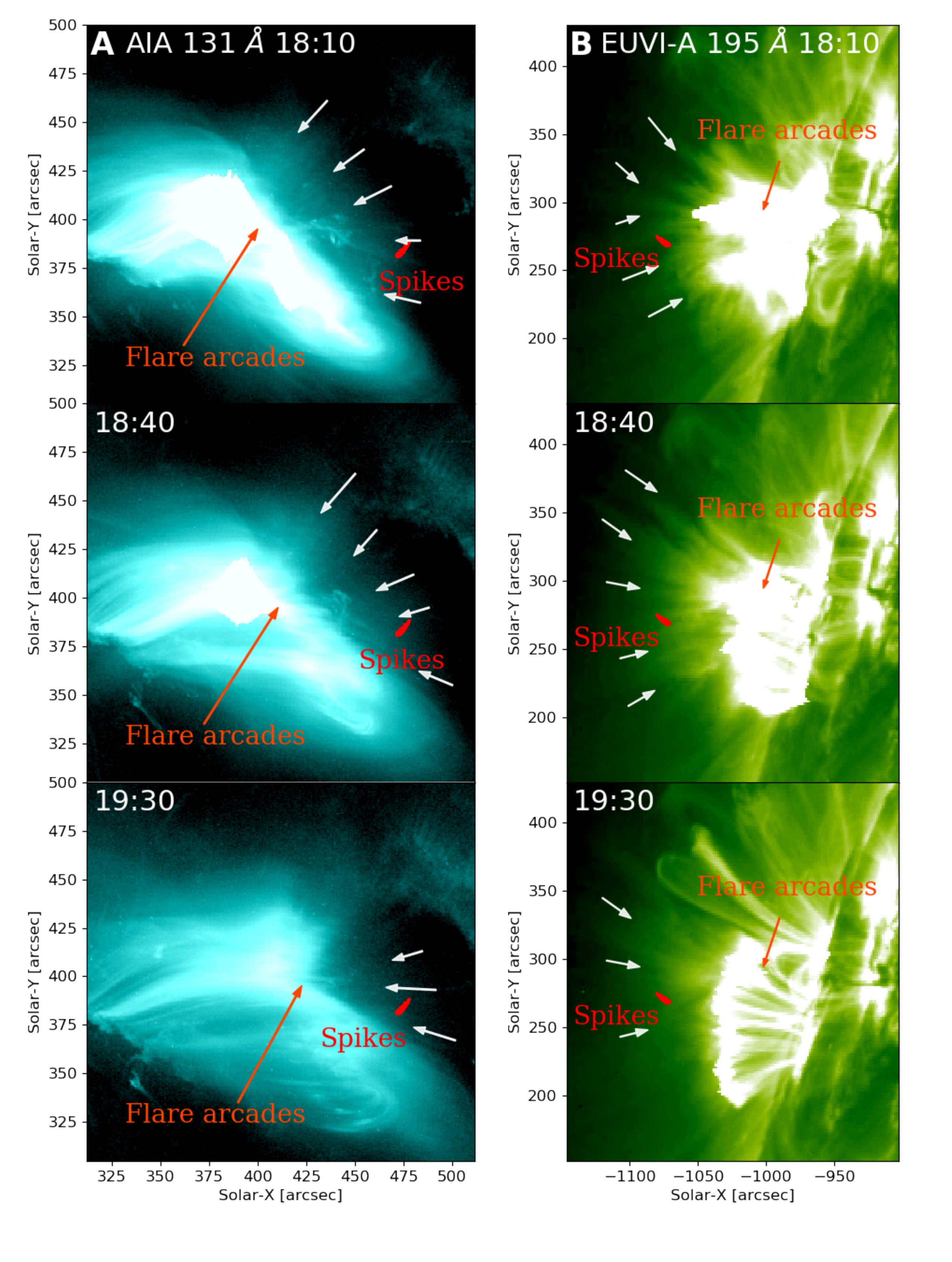}
\epsscale{1.2}
\caption{SAF structure above the flare arcades. The SAF structure (marked as white arrows) is observed in both \sdo/AIA 131 \AA\ (panel (a)) and \sta/EUVI 195 \AA\ (panel (b)) images from $\sim$17:50 to $\sim$19:30 UT. The spike burst source at $\sim$18:07 UT is also shown. The location of the spike source in \sta/EUVI's perspective is based on 3D reconstruction results discussed in Section \ref{sec:spikes_3d}. An animation of the dynamic SAF structure viewed from the two perspectives from 17:50 UT to 19:30 UT is available online. The duration of the animation is 3 s.\label{fig:saf}}
\end{figure}
\begin{figure*}
\plotone{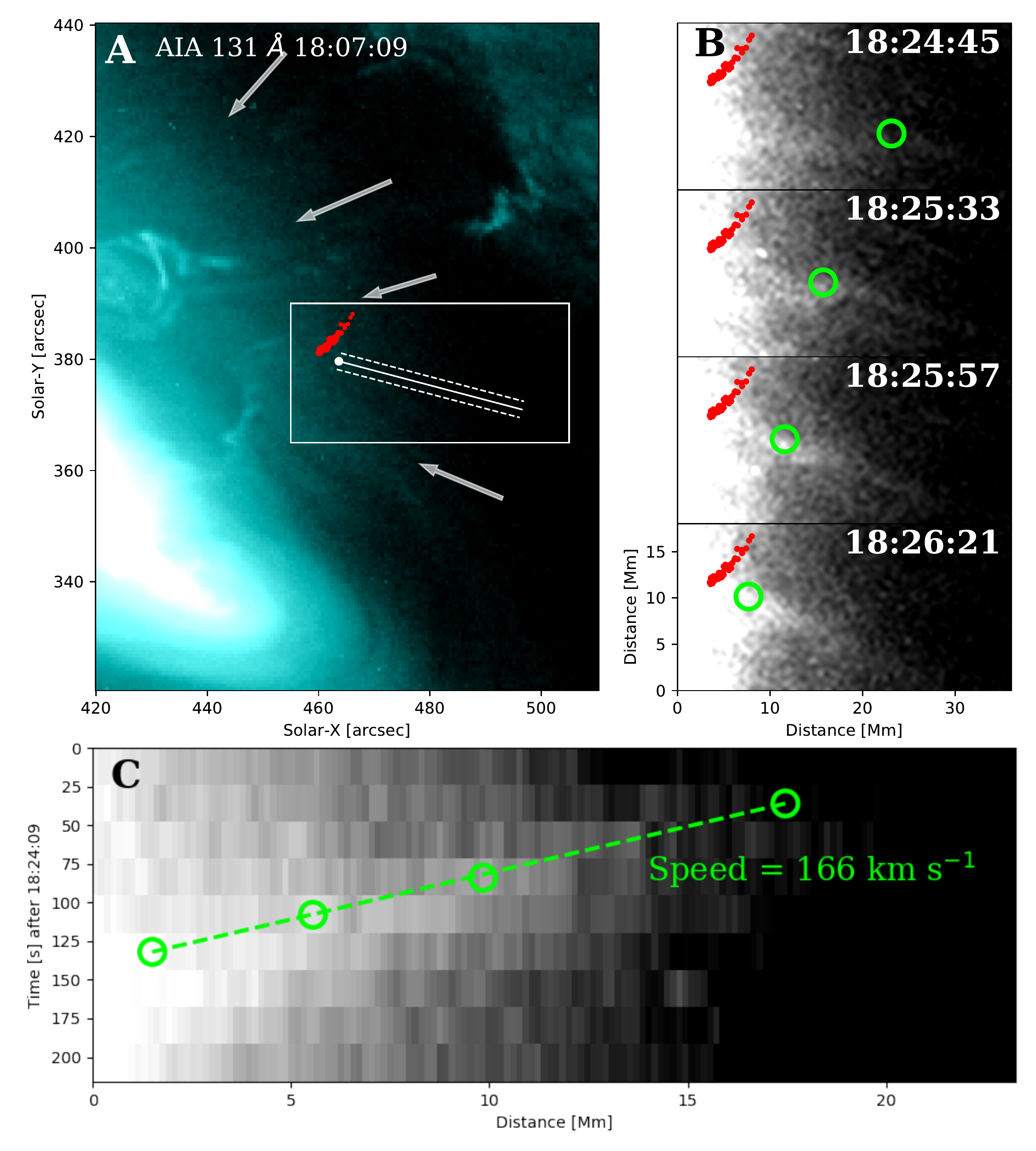}
\epsscale{1.0}
\caption{Example of plasma downflows observed in \sdo/AIA 131 \AA\ images. (a) Spike burst centroids at 18:07:07 UT (same as those in Fig. \ref{fig:radio}(e)) overlaid on AIA 131 \AA. The diffuse SAF structure is indicated by white arrows. (b) Example plasma downflow observed in the vicinity of the spike burst at $\sim$18:25 UT. The field of view of the AIA 131 \AA\ time-series images is indicated by a white box in panel (a). The images are overexposed to reveal the faint downflow feature. Green circles mark the leading front of the downflow. (c) Time--distance plot of the plasma downflow along the cut marked as a white line in panel (a) (the distance starts from the location of the white dot and increases westward, or toward the right-hand side). The speed of the downflow in projection is about 166 km s$^{-1}$. An animation of the downflow from 18:23:57 UT to 18:29:33 UT is available online. The duration of the animation is 1 s.\label{fig:sads_aia}}
\end{figure*}

In the previous subsection, we have established that the spike source is likely located well above the flare arcades visible in radio, EUV, and SXR wavelengths and, possibly, just above freshly reconnected magnetic loops, which is consistent with the flare termination shock interpretation. Here we discuss another important piece of evidence that further supports this scenario: the presence of an SAF and supra-arcade plasma downflows in the close vicinity of the spike source (or shock front). As shown in Figure~\ref{fig:saf}, in \sdo/AIA 131 \AA\ (panel (a)) and \sta/EUVI 195 \AA\ (panel (b)) time-series images (both have sensitivity to 10--20 MK hot plasma; \citealt{2008SSRv..136...67H,2010A&A...521A..21O}), a long-lasting SAF structure is seen above the flare arcades from $\sim$17:50 UT to $\sim$19:30 UT. Similar to those previously reported in other eruptive events \citep{2002SoPh..210..341G,2014ApJ...796...27I,2014ApJ...786...95H,2017ApJ...836...55R, 2020ApJ...905..165R,2019MNRAS.489.3183C}, such SAF structures often have a diffuse appearance with many finger-shaped fine structures. Within the SAF structure, multitudes of plasma downflows are identified at about 20 minutes after the time of the spike bursts. Figure~\ref{fig:sads_aia}(b) shows a clear example of a plasma downflow in \sdo/AIA 131 \AA\ time-series images that appeared at around 18:25 UT. In the time--distance map of Figure \ref{fig:sads_aia}(c), the speed of the downflow is found to be $\sim$166 km~s$^{-1}$ in projection, or $\sim$227 km~s$^{-1}$ if we adopt the same projection angle of the loop-top GS arcade $\varphi\approx43^{\circ}$.

\begin{figure}
\plotone{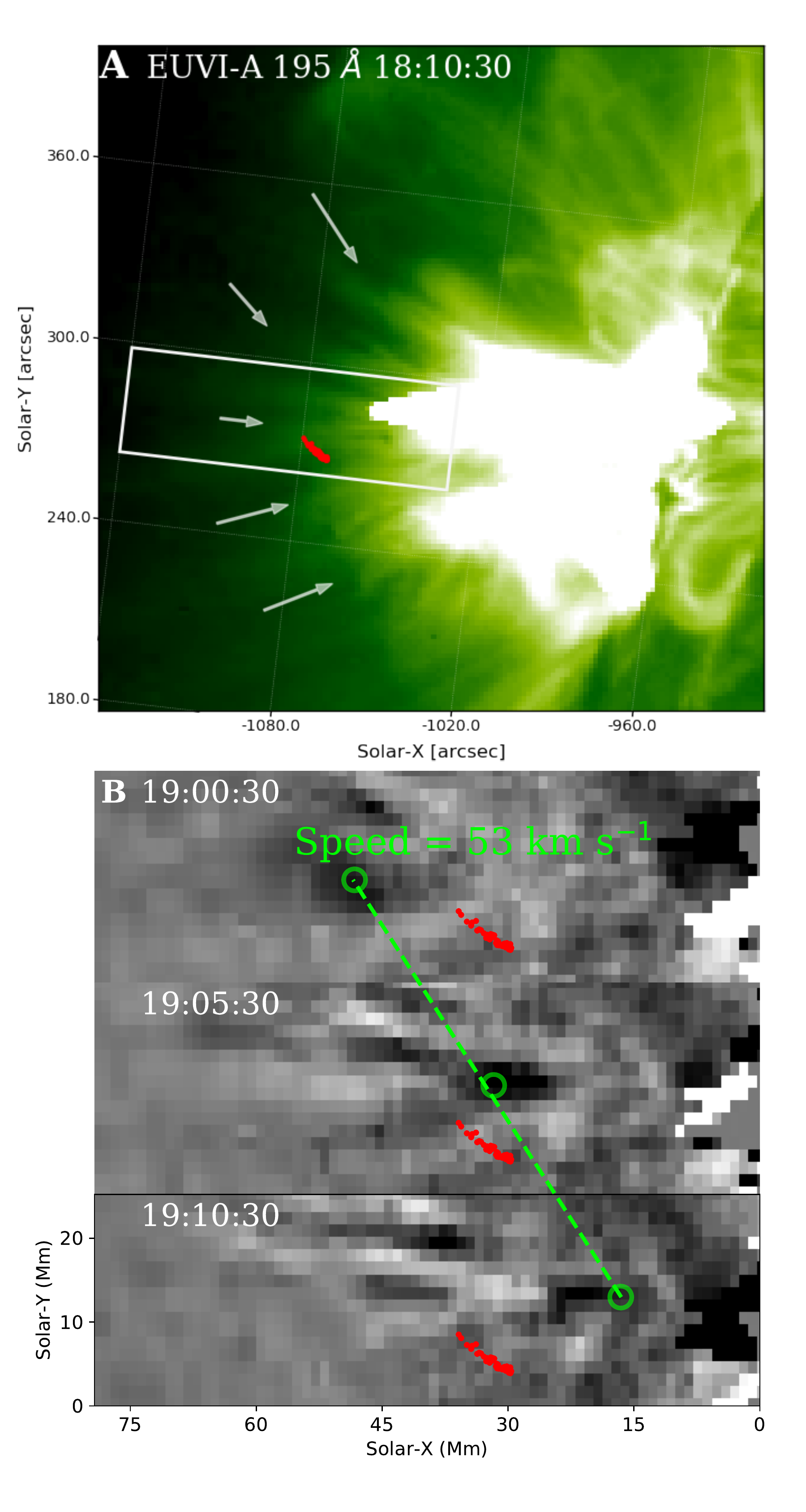}
\epsscale{1.0}
\caption{Example of plasma downflows observed in \sta/EUVI 195 \AA\ images. (a) Spike burst location at 18:07:07 UT (same as those in Fig. \ref{fig:saf}(b)) overlaid on EUVI 195 \AA. The diffuse SAF structure is indicated by white arrows. (b) Example plasma downflow observed in the vicinity of the spike burst at 19:00--19:10 UT. The field of view is indicated by a white box in panel (a). To show the downflow feature more clearly, running-ratio images are used. Note that although the downflows are tracked using dark moving features (green circle) in the running-ratio images, similar to those in AIA 131 \AA, they are in fact bright moving features in the original images. The speed in projection of this downflow is $\sim$53 km~s$^{-1}$. An animation of the example downflow from 18:35:30 UT to 19:20:30 UT is available online. The duration of the animation is 2 s. \label{fig:sads_sta}}
\end{figure}

Such an SAF structure is also clearly seen by \sta/EUVI 195 \AA\ with a limb-view perspective (Figure \ref{fig:saf}(b)). Based on our estimate of the 3D spatial location of the spike source discussed in the previous subsection, in Figure~\ref{fig:saf}(b), we show the inferred location of the same spike source in the \sta/EUVI 195 \AA\ images by applying the same projection angle $\varphi\approx 43^\circ$ for each spike centroid obtained at different frequencies. Similar to the spike location in \sdo/AIA's view, the spike source is also located close to the top of the SAF structure. Several plasma downflow events are also identified in \sta/EUVI 195 \AA\ running-ratio images (intensity ratio of the current frame to the previous frame 5 minutes earlier) later into the event. Figure~\ref{fig:sads_sta}(b) shows an example of such a plasma downflow event at 19:00 UT. The apparent speed of the downflow is $\sim$53 km s$^{-1}$ in projection. Such a speed appears much slower than those measured in \sdo/AIA 131 \AA\ images. However, we note that the low speed is most likely due to the selection bias: the low cadence of \sta/EUVI 195 \AA\ images (5 minutes, compared to 12 s of \sdo/AIA) only allows downflows with sufficiently slow speeds to be detected in consecutive frames.

\section{Discussions and Conclusions} \label{sec:discussion}

\begin{figure*}
\plotone{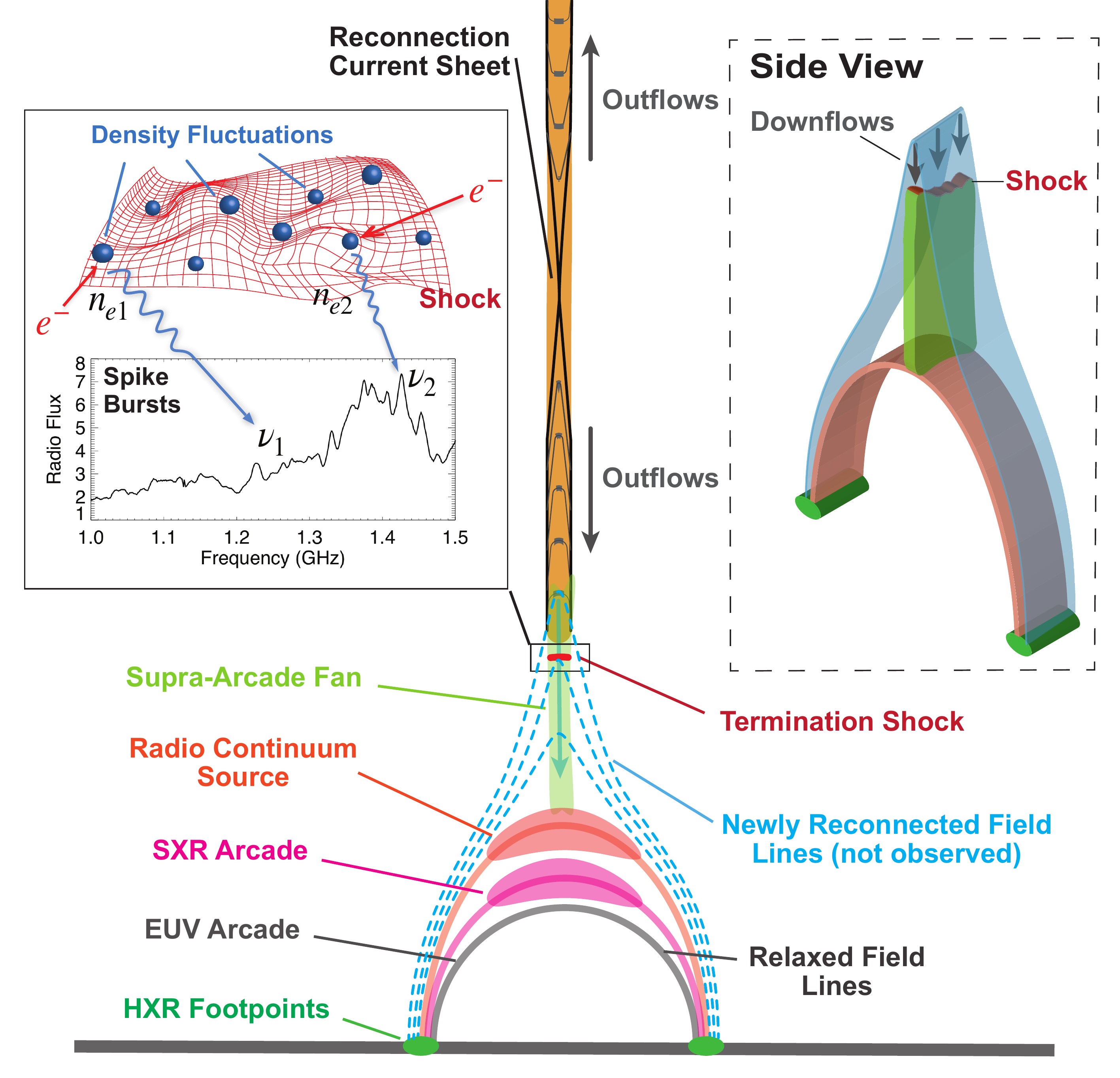}
\caption{Schematic illustration of observed phenomena during the extended energy release phase of the eruptive flare event. A filament eruption induces magnetic reconnection in a large-scale current sheet. Downward-propagating reconnection outflows impinge on the newly reconnected flare loops, forming a termination shock within the SAF region where highly bent, newly reconnected magnetic field lines are formed. The shock and the SAF are located above the radio continuum source (orange curve), SXR arcade (magenta curve), and EUV arcade (gray curve). The observed radio spike bursts are produced at the shock front where nonthermal electrons interact with density fluctuations (inset on the left; after Chen15 and Chen19). Accelerated electrons propagating downward along the flare arcade produce the conjugate HXR footpoints and (E)UV flare ribbons. The inset on the right shows a side view of the lower portion of the schematic.  
\label{fig:cartoon}}
\end{figure*}

Here we have presented another stochastic spike burst event recorded by the VLA with radio dynamic imaging spectroscopy. The event was observed during the extended energy release phase of the SOL2012-03-10 M8.4 eruptive flare event, which is much stronger than the SOL2012-03-03 C1.9 event studied by Chen15. In VLA's vector dynamic spectrum, the spike bursts show very similar spectrotemporal features to those reported in Chen15, which consist of myriad highly polarized, narrowband and short-lived spike bursts with a high radio brightness temperature of $10^9$ K. VLA's dynamic spectroscopy imaging technique allows each frequency and time pixel in the dynamic spectrum to be imaged and the corresponding source centroid to be located with high accuracy ($<$1''). The centroids of the spike bursts at different frequencies delineate a nearly linear feature, located well above the bright EUV/X-ray flare arcade in \sdo/AIA's view against the disk. 

With the aid of EUV observations from \sta/EUVI from a limb-view perspective, we reconstruct the 3D geometry of the filament eruption and the underlying flare loops observed in EUV, SXR, and radio wavelengths. By assuming that the spike source makes a similar projection angle to the newly formed flare arcade where a radio gyrosynchrotron source is present, we reconstruct the 3D location of the spike source and place it well above the flare loops. In both \sdo/AIA and \sta/EUVI images, we identify SAF-like structure in the close vicinity of the spike burst source, where multitudes of plasma downflows also occur.
\edit1{The speeds of the observed plasma downflows ($\sim$227 km s$^{-1}$ after deprojection) may be super-Alfv\'{e}nic only if the magnetic field in the spike source region is sufficiently low ($\lesssim$15 G, if we take a plasma density of $2\times 10^{10}\ \mathrm{cm}^{-3}$ for fundamental plasma radiation at 1.3 GHz). We also note that the downflows are only detected in AIA images $\sim$18 minutes after the spike bursts when the supra-arcade structures are more discernible; hence, it is possible that the downflows have already slowed down substantially. If the plasma downflows at the time of the spike bursts are indeed sub-Alfv\'{e}nic, it remains an outstanding question whether the observed plasma flows are indeed signatures of the reconnection outflows themselves that drive the termination shock or, instead, are structures in the outflows that are substantially slower owing to, e.g., an aerodynamic drag force \citep{2011ApJ...730...98S,2018ApJ...868..148L,2020ApJ...900...17Y}.}

Based on the observational evidence discussed above, we conclude that our observations are generally consistent with the scenario proposed by Chen15: a flare termination shock is formed at the front of reconnection outflows above the flare arcade. The observed spike bursts are due to coherent radio emission from nonthermal electrons in \edit1{either the upstream or downstream region} of the shock front. The schematic illustration in Fig. \ref{fig:cartoon} summarizes our observations of a variety of features in the flare reconnection picture during the extended energy release phase. However, unlike the limb event in Chen15, this event is viewed against the disk. Although we have demonstrated that the spike source is located well above the flare loops in an SAF region, the exact relationship between the spike source and the flare geometry is unclear. This is largely due to the unfavorable viewing geometry for seeing the newly reconnected, presumably cusp-shaped field lines between the shock front and the flare loops (cyan curves in Figure \ref{fig:cartoon}). On the other hand, the cooling flare loops seen 2 hr later below the spike burst location are a possible indication for the presence of these newly reconnected field lines.

Another \edit1{intriguing} difference is that spatial separation between the spike source and the loop-top SXR/radio sources in this event ($\sim$70$''$ in projection and similar in radial height if we adopt a projection angle of $43^{\circ}$) is much larger than the $\sim$8$''$ separation between the spike source and the HXR loop-top source reported in Chen15. Theoretical studies of the flare termination shock suggest that the shock occurs above the flare arcades where there is a sharp gradient in flow speed---i.e., where the supermagnetosonic reconnection outflows are ``terminated'' as they impinge on the dense, previously reconnected plasma above the loop tops. The exact location of the shock front, however, depends on a variety of factors, including the flare geometry, properties of the reconnection diffusion region, and flow speeds in the shock upstream and downstream \citep{1986ApJ...305..553F}. As shown by recent numerical results in \citet{2019MNRAS.489.3183C}, early in the flare the shock occurs at a relatively low position above the loop tops. At later times into the flare, after more and more reconnected magnetic flux (and plasma) piles up in the loop-top region, the distance between the shock front and the loop top appears to grow significantly (see their Figure~14(f)). For our event, as the spike bursts are observed well into the extended energy release phase ($\sim$25 minutes after the SXR peak), it is possible that the termination shock front is located at a substantial distance away from the top of the flare arcades, similar to those reported in earlier works when the radio signature was detected late in the flare \citep[e.g.,][]{2004ApJ...615..526A}. 

Finally, we note that there are several unanswered questions in the interpretation of the observations in terms of the flare termination shock scenario. One question is on why the stochastic spike bursts \edit1{do not usually coincide with the peak of the flare energy release and, in fact, are only occasionally reported within a limited period \citep[see also][]{2002A&A...384..273A,2004ApJ...615..526A}.} We attribute the rather uncommon appearance of these spikes to (a) the requirements for generating a fast-mode flare termination shock, which may only be met in certain locations and times \edit1{(not necessarily the impulsive energy release phase)} in the loop-top region, where the outflow speed exceeds the local fast-mode magnetosonic speed \citep[e.g.,][]{2018ApJ...869..116S}, and (b) the special conditions for producing the coherent radio emission. Chen15 suggests a linear mode conversion mechanism due to Langmuir waves interacting with small-scale density fluctuations. The mechanism requires (1) the excitation and nonlinear growth of Langmuir waves and (2) the subsequent interactions with the turbulent medium. Both conditions are not universally fulfilled throughout the shock region. The second intriguing question is on the requirement of a localized dense region in the spike source that has a plasma density of order $10^{10}$ cm$^{-3}$ (suggested by the plasma radiation mechanism). \edit1{Although we argue that such a plasma density in the above-the-loop-top region is not inconsistent with the flare context (Appendix A), it may require some additional localized density enhancement or compression.} The lack of a strong EUV emission in this region suggests that each of the density fluctuations on the shock surface may be too compact to have a sufficiently large differential emission measure so as to be distinguished against the coronal background (see, e.g., \citealt{2013ApJ...763L..21C} for similar discussions on the absence of type-III-radio-burst-emitting coronal loops in EUV images). Another question is on the apparent frequency gradient along the shock surface (see Figure \ref{fig:radio}(e)), which presumably corresponds to a localized density gradient. Such a density gradient may be attributed to the spatiotemporal variations of the shock properties along the shock front \citep{2019ApJ...884...63C}. Last but not least, as discussed earlier, the seemingly sub-Alfv\'{e}nic plasma downflows (albeit observed at much later times) and the absence of strong modulations of the spike-centroid-illuminated shock front, such as those reported in Chen15, both call for further investigation. For the latter, based on MHD simulations, \citet{2018ApJ...869..116S} suggested that the shock dynamics is intimately related to the details of the plasma-outflow--shock front interactions. 

In order to answer the questions above and gain a more comprehensive understanding on the formation of the flare termination shocks and their associated radio emission, more observations made with radio dynamic spectroscopic imaging are required. Complementary high-resolution, high-cadence EUV/X-ray imaging and spectroscopic observations, as well as 3D numerical simulations, are also particularly helpful for understanding the essential physical context in the vicinity of the shock, including SAF structures, plasma outflows, and the hot flare arcades.

\begin{figure}
\plotone{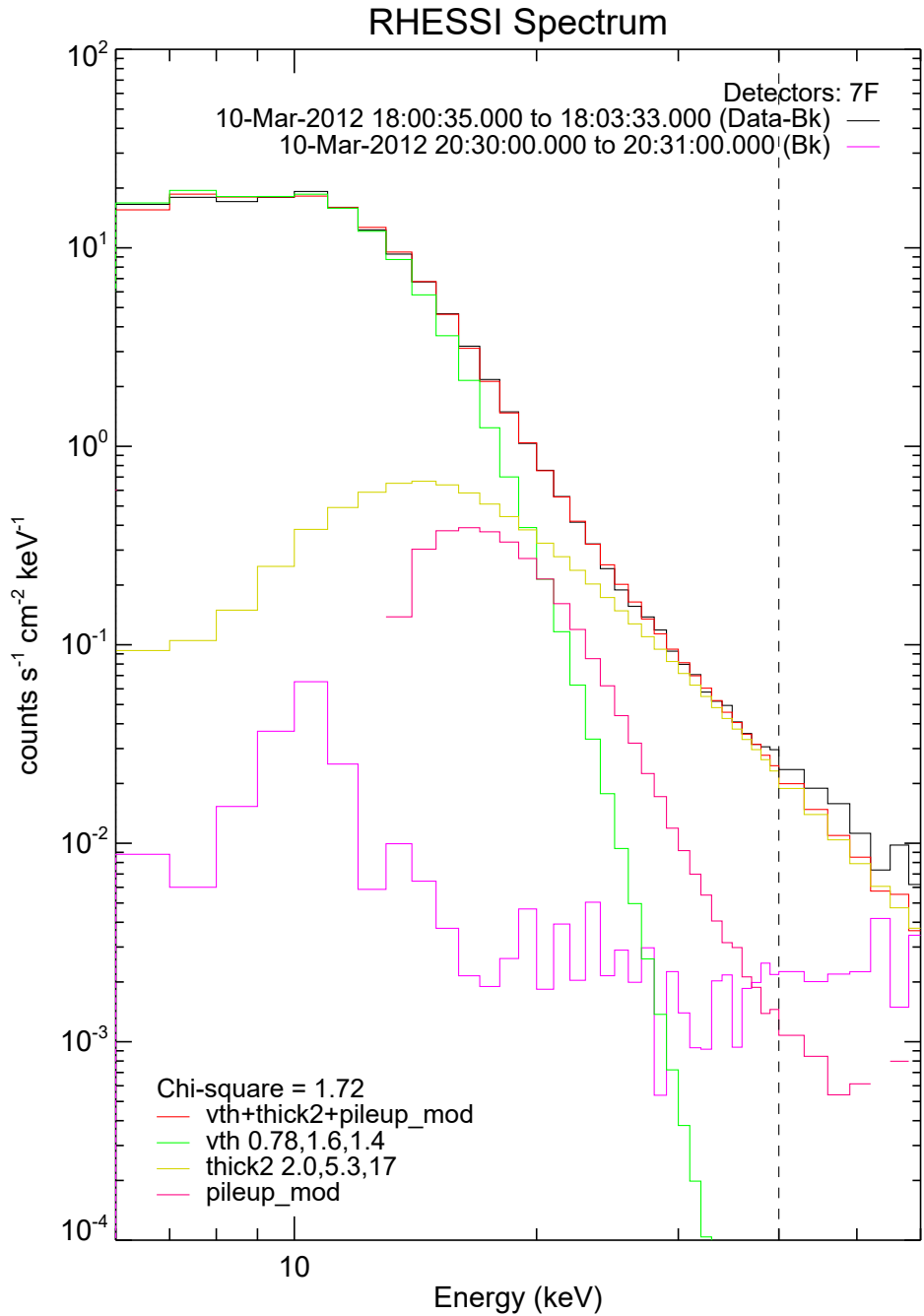}
\epsscale{0.6}
\caption{RHESSI X-ray spectral analysis of detector 7 for time interval 2 (18:00:35 UT--18:03:33 UT, as marked in Figure~\ref{fig:lightcurve}(a)). For the thermal component (``vth''; green curve), three fit parameters (indicated in the legend) are volume emission measure ($\textrm{EM}_V = 7.8\times10^{48}$ cm$^{-3}$), plasma temperature (1.6 keV, or 19 MK), and relative abundance (1.4). For the thick-target model, we adopt a single power-law distribution for the nonthermal electrons. Three fit parameters are total integrated electron flux ($2.0\times10^{35}$ s$^{-1}$), power-law index of the electron flux spectrum ($\delta = 5.3$), and low-energy cutoff (17 keV). The high-energy cutoff is fixed to 1 MeV. A pileup component is also included in the spectral fit. The upper limit of the fit is restricted to 40 keV to avoid the possible contribution of the background at higher energies. See the Appendix for details. \label{fig:rhefit}}
\end{figure}

\edit1{
\appendix
\section{X-Ray spectral Analysis}\label{appendixa}
We perform RHESSI X-ray spectral analysis based on data from 18:00:35 UT to 18:03:33 UT (time  interval 2) using front segments of detectors 1, 3, 5, 6, 7, 8, and 9 separately. The attenuator state within the selected time interval is A1. A pileup check is performed with \texttt{hsi\_pileup\_check.pro} in \texttt{SSWIDL}, which suggests a small but nonnegligible pileup effect. The background time is selected as 20:30:00--20:31:00 UT (during RHESSI night; purple curve in Figure~\ref{fig:rhefit}). By examining the spectra from various detectors, we suspect that there could be additional background contribution to the counts above $\sim$40 keV that is unaccounted for by the nominal background derived from the selected interval. Therefore, we restrict the fit range to 6--40 keV.}  

\edit1{
For the spectral fit, we adopt an isothermal model for the thermal plasma, a thick-target model with a single power-law distribution for the nonthermal electrons, as well as a pileup component. The fit results and parameters from seven different detectors are generally consistent with each other. An example from detector 7 is shown in Figure~\ref{fig:rhefit}. For the thermal component contributed by the SXR flare arcade (green curve), the average plasma temperature across the seven detectors is 19 MK (the full range is 18--20 MK) and the average volume emission measure is $\mathrm{EM_V}\approx7.8\times10^{48}$ cm$^{-3}$ (the full range is $(6.9-9.1)\times10^{48}$ cm$^{-3}$). 
We take the 50\% contour of the 12--15 keV SXR flare arcade source to estimate a source area of $A \approx 38''\times19'' \approx 3.8\times10^{18}\ \mathrm{cm}^{-3}$. If we take the column depth of the SXR arcade to be the same as its width, we estimate a volume of $V\approx5.3\times10^{27}\ \mathrm{cm}^{3}$. 
The thermal density is hence $n_{\rm th}=(\mathrm{EM}_V/V)^{1/2}\approx3.8\times10^{10}\ \mathrm{cm}^{-3}$, which is about two times the inferred plasma density of the spike burst source for fundamental plasma radiation at 1.3 GHz ($2\times10^{10}\ \mathrm{cm}^{-3}$). We conclude that the density of the spike source is not inconsistent with the flare context, particularly if a large density scale height (e.g., a typical flare temperature of 10 MK corresponds to a hydrostatic scale height of $\sim$500 Mm) and/or a localized density enhancement or compression are present at the spike burst site. } 

\acknowledgments

This work makes use of public VLA data from the observing program VLA/11B-129. The authors acknowledge Richard Perley, Michael Rupen, Ken Sowinski, and Stephen White for their help in carrying out the observing program. The NRAO is a facility of the National Science Foundation (NSF) operated under cooperative agreement by Associated Universities, Inc. This work is supported partly by NASA grant NNX17AB82G and NSF grants AGS-1654382, AGS-1723436, and AST-1735405 to the New Jersey Institute of Technology. S.K. was
supported by NASA contract NAS 5-98033 for RHESSI. We thank Lucia Kleint and Marina Battaglia for providing us the original version of the 3D ellipse loop fitting codes. We also thank Chengcai Shen and Kathy Reeves for helpful discussions.

\facilities{VLA, SDO, STEREO, RHESSI}

\software{CASA \citep{2007ASPC..376..127M},
          Astropy \citep{2018AJ....156..123A}, 
          SunPy \citep{sunpy_community2020}.
          }

\bibliography{references}

\end{document}